\documentclass[a4paper,twoside,english,amsfonts,amssymb,amsmath,preprint,showkeys,nofootinbib]{revtex4-1}
\usepackage[utf8]{inputenc}
\usepackage{color}
\usepackage{babel}
\usepackage{array}
\usepackage{multirow}
\usepackage{amstext}
\usepackage{graphicx}
\usepackage{rotating}
\usepackage[unicode=true,
 bookmarks=false,
 breaklinks=false,pdfborder={0 0 1},backref=section,colorlinks=false]
 {hyperref}
\hypersetup{pdftitle={Article},
 pdfauthor={Author},
 pdftex}

\makeatletter

\providecommand{\tabularnewline}{\\}

\makeatother

\begin{document}
\title{Investigating the Single Production of Vector-Like Quarks Decaying
into Top Quark and W Boson through Hadronic Channels at the HL-LHC}
\author{A.C. Canbay}
\email[Correspondence email address: ]{ acanbay@cern.ch}

\affiliation{Department of Physics, Ankara University, 06100, Ankara,Turkiye}
\author{O. Cakir}
\email[Correspondence email address:]{ ocakir@science.ankara.edu.tr}

\affiliation{Department of Physics, Ankara University, 06100, Ankara,Turkiye}
\date{\today}
\begin{abstract}
We investigate the single production of vector-like quarks at the
High Luminosity LHC (HL-LHC). With the assumed (enhanced) couplings
to third generation quarks of the standard model, vector-like quarks
$B/X$ are produced in association with a bottom ($b$) or top ($t$)
quark, which correspond to $Bbq$ and $Btq/Xtq$ production modes,
including an additional soft forward jet from the spectator quark
($q$). This study focuses on high-mass vector-like quarks $B/X$
decaying into a top quark and a $W$ boson, resulting in the final
state jets emerging from hadronically decaying top quark ($t\to Wb$)
and $W$ boson ($W\to q\bar{q}'$). The events with $W$ boson and
$t$ quark have been analysed using tagging techniques for large-radius
jets. The scan ranges of the mass ($1000<m_{B/X}<3000$ GeV) for the
relative width $\Gamma_{B/X}/m_{B/X}=0.1$ and $\Gamma_{B/X}/m_{B/X}=0.01$
of vector-like $B/X$ quarks have been investigated. From the results
of the analysis, the masses of vector like quarks $B$($X$) up to
$2491$ ($2364$) GeV and $2018$ ($1873$) GeV can be excluded corresponding
to these relative width cases at $95\%$ CL depending on the type
and branching scenarios at integrated luminosity projection of $3$
ab$^{-1}$ at the HL-LHC.
\end{abstract}
\maketitle

\section{Introduction}

\label{sec:outline-1}

The most of the phenomenological studies are focused on exploiting
experimental data from the previous runs of particle colliders which
provide the data corresponding to a luminosity delivered to particle
detectors. A precise total cross section and differential cross section
measurements of the single top quark production have been performed
using data at $\sqrt{s}=13$ TeV by the ATLAS and CMS Colllaborations
\citep{ATLAS2018,CMS2018}. These studies aim also to pose constraints
on theoretical models beyond the standard model (SM) of particle physics.
These are also trying to predict the exclusion and/or discovery reaches
of new searches during the next runs.

A variety of extensions of the SM predict the existence of new heavy
particles. New heavy quarks (heavier than top quark) are generally
expected to be of vector-like nature if they exist. These particles
could have a role in the stabilization of the Higgs boson mass, and
hence promote a potential solution to the hierarchy problem. Vector-like
quarks (VLQs) are color triplets and their left-handed and right-handed
components transform in the same way \citep{Aguilar-Saavedra2009}
under the electroweak symmetry group $SU(2)_{L}\times U(1)_{Y}$ of
the SM.

In the predicting models, vector-like quarks are expected to couple
mostly to third generation quarks \citep{Vignaroli2012,Aguilar-Saavedra2013a,Aguilar-Saavedra2013b},
and they can have both charged and neutral current interactions. A
down type vector-like $B$ (VLB) quark with $-1/3$ can decay into
$Wt$, $Zb$ or $Hb$ (their charge conjugation can also take place),
while an up-type vector-like T (VLT) quark can decay into $Wb$, $Zt$
or $Ht$ (and similarly their charge conjugation). The VLQs can arise
in multiplets, such as singlets, doublets or triplets. In the minimal
models, each scenario results in different $T$ and $B$ branching
ratios. For singlets, the branching ratios are $50\%$ for $B\to Wt$
and $T\to Wb$, and $25\%$ for $B\to Hb/Zb$ and $T\to Ht/Zt$. A
roadmap for the vector-like singlet quark search has been reviewed
theoretically and phenomenologically in Ref. \citep{Alves2023}. However,
in one of the doublet scenario the $B$ decays only to $Hb$ and $Zb$
with equal branchings ratios of $50\%$, and similarly the $T$ decays
only to $Ht$ and $Zt$ with equal branchings, in another doublet
scenario the $B$ decays only to $Wt$, and similarly the $T$ decays
only to $Wb$ with $100\%$ branching ratio \citep{Aguilar-Saavedra2013a,Aguilar-Saavedra2013b}.
Experimental searches at the LHC mainly focused on singlets $(T)$
and $(B)$ with charges of $2/3$ and $-1/3$, respectively; or doublets
$(X\;T)$, $(T\;B)$ and $(B\;Y)$ with the left and right chiralities
for each of them. Here, vector-like $X$ (VLX) and $Y$ (VLY) have
their exotic charges of $5/3$ and $-4/3$, respectively. These can
couple to the SM quarks through only charged currents, leading to
decays $X\to Wt$ and $Y\to Wb$ with a branching ratio of $100\%$. 

The model-independent pair production (via strong interaction) search
results from the ATLAS and CMS collaborations, which set a limit on
VLQ masses in the range of O(1 TeV) independently of its electroweak
representation. The single production of vector-like quarks (via electroweak
interactions) is affected by both the coupling strengths to the SM
quarks and their masses, where a jet is emitted at a low angle to
the beam direction. However, depending on multiplet structure, mass
and coupling strength of VLQs the single production may overcome their
pair production above few TeV range. 

Previously, the ATLAS Collaboration have searched for the production
of single vector-like quarks in the Wt final state in pp collisions
at $\sqrt{s}=8$ TeV and set limits on the cross section of the single
production of VLB quarks decaying into the Wt final state using a
novel approach for boosted event topologies \citep{ATLAS2016}. Searches
performed recently by ATLAS and CMS Collaborations set limits on masses
and couplings of different type of VLQs using proton-proton collision
data at a center of mass energy of $13$ TeV using Run 2 recorded
data and simulated signal and background samples. The all hadronic
final state is used for single vector-like B quark production, and
searched in CMS \citep{CMSCollab2018a} in $Hb$ intermediate states
leading to three $b$-tag jets, two of them are reclustered in a Higgs-tag
jet. The signature of the single production of VLB is used to categorize
the signal region and separate it from the background. The search
has used an integrated luminosity $35.9$ fb$^{-1}$ of data and set
limits on the masses of $920$ to $1490$ GeV for vector-like quarks
$B/X$ in the single lepton mode, at a relative width of $10\%$ \citep{CMS2019}.
Recently, the search is carried out on 139 fb$^{-1}$ of proton--proton
collision data at $\sqrt{s}=13$ TeV collected with the ATLAS detector
between 2015 and 2018 runs. This search excludes the presence of a
vector-like B quark in the full hadronic mode with mass ranging between
$1.0$ TeV and $2.0$ TeV for coupling $\kappa=0.3$ \citep{ATLASCollab2021a}.
Currently, some results are using the full Run 2 statistics, and more
results are expected to provide the total available statistics very
soon. In extension, the projections can be made for the High-Luminosity
LHC (HL-LHC). 

High energy particle collisions at the HL-LHC can lead to the production
of massive particles (e.g. $W/Z/H$ bosons and top quarks) with much
larger transverse momentum ($p_{T}$) than their rest mass. The decay
products of such particles tend to be collimated, or 'boosted' along
the direction of the parent particle. If the massive particles are
sufficiently boosted, their overlapping hadronic decay products cannot
be well reconstructed with small radius jets, and require large radius
(large-$R$) jet reconstruction. The identification of hadronically
boosted $W$ boson decays with large-R jets is vital in many physics
analyses at the LHC and HL-LHC. The dominant backgrounds are the jets
originating from light quarks and gluons (QCD jets) as they occur
at a much higher rate than boosted $W$-jets at high energy hadron
collisions. Thus, boosted boson tagging and top tagging are used as
the key techniques for searches of VLB/VLX in all hadronic decay mode,
which also suppress the relevant background efficiently.

This paper is organized as follows: Section II describes the signal
and background event generation and the simulation samples that are
used in the physics analysis. We make estimations for the single production
of vector-like quarks at the HL-LHC at an integrated luminosity projection
of $3$ ab$^{-1}$. With the assumption of enhanced couplings to third
generation quarks of the standard model, vector-like $B$ or $X$
quark is produced in association with a bottom ($b$) or top ($t$)
quark, which lead to $Bbq$ and $Btq/Xtq$ production modes, including
an additional forward jet from the spectator quark ($q$). This study
focuses on the intermediate state objects top quark and $W$ boson
(through $B\to Wt$ decay), and final state jets from hadronically
decaying top quark ($t\to Wb$) and $W$ boson ($W\to q\bar{q}'$).
Section III describes the object definition, event selection, tagging
and analysis of the simulated events. Section IV summarizes the reconstruction
and statistical significance of VLB/VLX quark signal including the
systematic uncertainties. Finally, in section V, the conclusions have
been drawn.

\section{Signal and Background Event Generation}

\label{sec:signal_background-1}

\begin{figure}[t]
\includegraphics[scale=0.4]{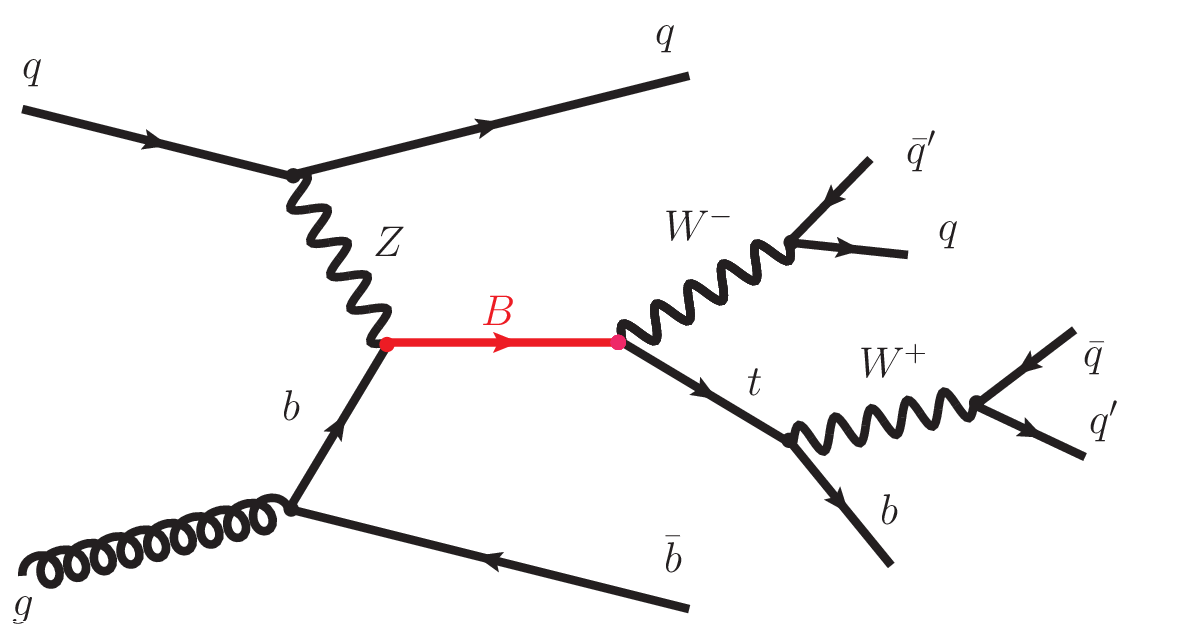} \includegraphics[scale=0.4]{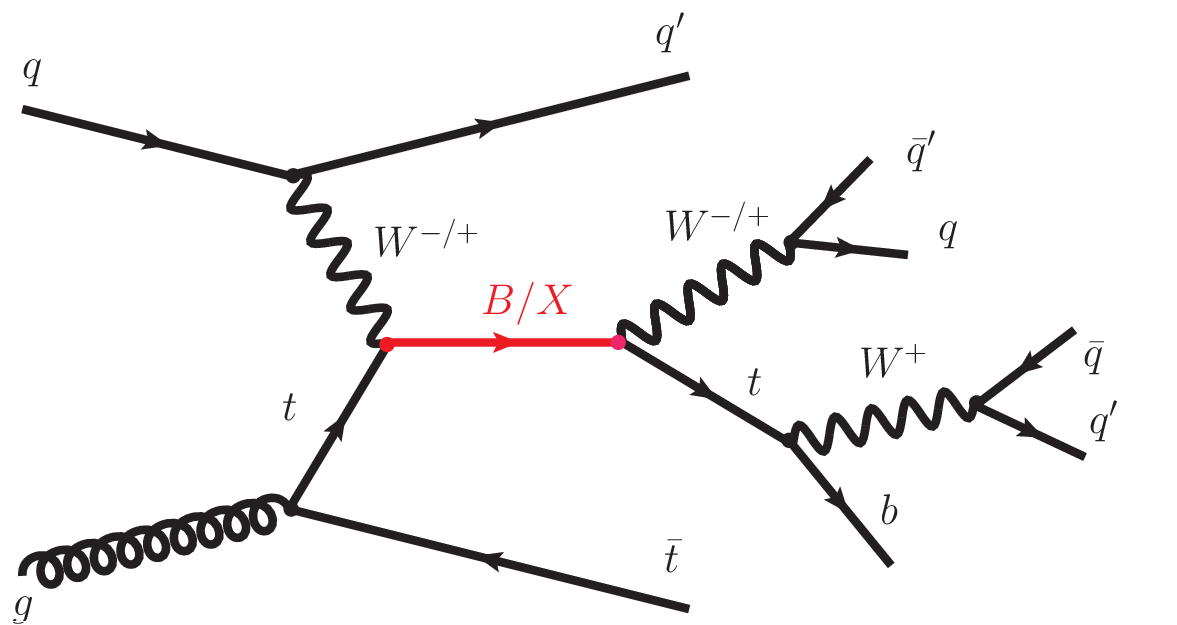}
\caption{Leading order representative diagrams for single production of vector-like
$B$ quark ($Z$ mediated (left) and of vector-like $B/X$ quark $W$
mediated (right)) in association with a bottom quark (left) or top
quark (right) and a light flavour quark, with the subsequent decays
to a top quark and $W$ boson leading to all hadronic mode.\label{fig1-1}}
\end{figure}

The leading order representative Feynman diagrams of single production
of VLB/VLX quark and decay chain are presented in Fig. \ref{fig1-1}.
Note that the analysis has also included the charge-conjugate process.
The VLB quark in this analysis is assumed to be in singlet or doublet
representation, while the VLX quark is assumed to belong to a doublet.
When the multiplet structure of vectorlike quarks is assumed, the
possible final states and branching ratios require an approach involving
simultaneous consideration of several final states \citep{Aguilar-Saavedra2013b}.
The study reported here significantly extend the sensitivity to events
in which a singly produced VLB/VLX quark decays to $Wt$ followed
by the hadronic decays $t\to Wb$ and $W\to jj$ in the resolved process.
The use of fully hadronic decays allows the direct reconstruction
of the VLB/VLX quark final state, and increases the expected signal-to-background
ratio in the signal region defined for the search. Its sensitivity
is considered to be studied by using tagging techniques resulting
in a signal-to-background improvement. 

This fully hadronic final state is of particular interest for vectorlike
quark masses above $1$ TeV. The resulting high-$p_{T}$ jets from
the top quark and $W$ boson are “boosted”, so that the decay products
of the top quark and $W$ boson are collimated and captured in two
large-radius (large-R) jets. This final state has the largest branching
fraction of all the potential $Wt$ decay modes and the large-R jets
can be identified as either $W$-boson or top-quark candidates through
tagging algorithms that use the substructure within the jet \citep{ATLAS2019}.
In addition, bottom-quark jet identification ($b$-tagging) provides
background rejection with high efficiency given the bottom-quark jet
coming from $t\to Wb$ decays. Assuming the existence of single VLB/VLX
quark production within the narrow width approximation (NWA) for the
parameter of relative decay width $\Gamma_{B/X}/m_{B/X}=0.1-0.01$,
the signal would appear as an excess of events with $Wt$ invariant
masses around the VLB/VLX quark mass. 

The model framework for the VLQs has been used from the UFO format
of VLQ model \citep{VLQmodel}. Signal and background events are generated
using event generator MadGraph5\_aMC-NLO \citep{MG5} with the parton
distribution function of NNPDF31 pdf set with \emph{lhapdfid} 324900
\citep{NNPDF31}, and the parton showering and hadronisation are performed
with Pythia8 \citep{Pythia8}. The signal cross sections for different
production modes ($pp\to Bbq$, $pp\to Btq/Xtq$) are given in Table
\ref{tab:Cross-sections of the signal}. We have used all hadronic
channels for each signal samples.

\begin{table}[h]
\centering \caption{Cross sections $\sigma(pp\to Bbj)$and $\sigma(pp\to Btj/Xtj)$ (in
pb) for $Bbq$ and $Btq/Xtq$ production process for different benchmark
mass points considered in the analysis, where the couplings are calculated
for a width to mass ratio of $\Gamma_{B/X}/M_{B/X}=0.1$ with the
branching ratios $BR(B\to Wt)=0.5$, $BR(B\to Zb)=0.25$ and $BR(B\to Hb)=0.25$
corresponding to a singlet state. The numbers in the paranthesis correspond
to $Xtj$ production related to the coupling $\kappa_{W}^{X}$ and
branching ratio $BR(X\to Wt)=1.0$. Cross section may depend on the
chirality of the vector like quarks when different representation
is considered. \label{tab:Cross-sections of the signal}}

\begin{tabular}{|c|c|c|c|c|c|}
\hline 
Mass {[}GeV{]} & $\kappa_{W}^{B}(\kappa_{W}^{X})$ & $\kappa_{Z}^{B}$ & $\kappa_{H}^{B}$ & $\sigma(pp\to Bbj)$ & $\sigma(pp\to Btj(Xtj))$\tabularnewline
\hline 
\hline 
$1000$ & $0.4019(0.5683)$ & $0.3840$ & $1.6105$ & $2.570\times10^{-1}$ & $1.032\times10^{-1}(2.075\times10^{-1})$\tabularnewline
\hline 
$1200$ & $0.3302(0.4670)$ & $0.3200$ & $1.6027$ & $9.837\times10^{-2}$ & $4.028\times10^{-2}(8.107\times10^{-2})$\tabularnewline
\hline 
$1400$ & $0.2807(0.3969)$ & $0.2743$ & $1.5981$ & $4.095\times10^{-2}$ & $1.711\times10^{-2}(3.447\times10^{-2})$\tabularnewline
\hline 
$1600$ & $0.2442(0.3454)$ & $0.2400$ & $1.5951$ & $1.866\times10^{-2}$ & $7.940\times10^{-2}(1.596\times10^{-2})$\tabularnewline
\hline 
$1800$ & $0.2163(0.3059)$ & $0.2133$ & $1.5930$ & $8.937\times10^{-3}$ & $3.879\times10^{-3}(7.809\times10^{-3})$\tabularnewline
\hline 
$2000$ & $0.1942(0.2746)$ & $0.1920$ & $1.5916$ & $4.440\times10^{-3}$ & $1.962\times10^{-3}(3.952\times10^{-3})$\tabularnewline
\hline 
$2200$ & $0.1762(0.2491)$ & $0.1745$ & $1.5905$ & $2.342\times10^{-3}$ & $1.038\times10^{-3}(2.086\times10^{-3})$\tabularnewline
\hline 
$2400$ & $0.1612(0.2280)$ & $0.1600$ & $1.5897$ & $1.248\times10^{-3}$ & $5.541\times10^{-4}(1.116\times10^{-3})$\tabularnewline
\hline 
$2600$ & $0.1487(0.2102)$ & $0.1477$ & $1.5890$ & $6.782\times10^{-4}$ & $3.055\times10^{-4}(6.139\times10^{-4})$\tabularnewline
\hline 
$2800$ & $0.1379(0.1950)$ & $0.1371$ & $1.5885$ & $3.790\times10^{-4}$ & $1.700\times10^{-4}(3.426\times10^{-4})$\tabularnewline
\hline 
$3000$ & $0.1286(0.1819)$ & $0.1280$ & $1.5881$ & $2.156\times10^{-4}$ & $9.684\times10^{-5}(1.947\times10^{-4})$\tabularnewline
\hline 
\end{tabular}
\end{table}

The cross sections of the relevant backgrounds and corresponding modes
are given in Table \ref{tab:The-cross-sections of the background}.
For the benchmark mass values of $1500$, $2000$ and $2500$ GeV,
the couplings for the corresponding modes are calculated as $0.083:0.081:0.505$
for $\kappa_{W}:\kappa_{Z}:\kappa_{H}$ at a relative decay width
of $\Gamma_{B}/m_{B}=0.01$. A single decay mode for VLX lead to coupling
values of $\kappa_{W}=0.117,$$0.087$ and $0.069$ for the same benchmark
mass values at $\Gamma_{X}/m_{X}=0.01$. We have generated events
by using the SM-full model of MadGraph5\_aMC-NLO \citep{MG5} for
top pair production, single top production and diboson production.
We have also generated $W+jets$ and $Z+jets$ backgrounds where a
matching and merging applied. 

The top quark pair production background samples are normalised to
their theory predictions. The $t\bar{t}$ predicted cross section
is $\sigma=599.0$ pb, the single top predicted cross sections are
$\sigma_{tj}=0.7394$ pb , $\sigma_{tb}=8.655$ pb and $\sigma_{Wt}=0.2553$
pb, which have been calculated at leading order. The event generator
is used to generate $W+jets$ and $Z+jets$ events. For the $W+nj$
and $Z+nj$ modes, where $n\leq4$, we have used $qcut=40$ GeV value
and $xqcut=20$ GeV value for matching scheme within MadGraph5\_aMC-NLO
\citep{MG5}. A parton jet matching scheme is employed to avoid double-counting
of partonic configurations generated by both the matrix-element calculation
and the parton shower \citep{MLM2002}. These samples are generated
separately for $W/Z$ with jets. The cross section for $W+nj$ where
$n\leq4$ is calculated as $\sigma_{Wj}=2.674\times10^{5}$ pb, and
$Z+nj$ where $n\leq4$ is calculated as $\sigma_{Zj}=8.219\times10^{4}$
pb. Diboson events ($WW$, $WZ$, $ZZ$) are generated with the same
event generator for the modelling of the underlying event, which undergo
further process of showering and hadronisation as explained before.
The corresponding cross sections for these processes are calculated
as $\sigma_{WW}=79.37$ pb , $\sigma_{WZ}=30.21$ pb and $\sigma_{ZZ}=11.56$
pb, respectively.

After event generation, the signal and all background samples are
passed through the fast simulation of the ATLAS detector \citep{ATLASCollab2008}
based on a modular framework Delphes \citep{DELPHES2014} and they
are reconstructed using the procedure for the simulation data to be
used in the analysis. For using the anti-kt jet algorithm \citep{Anti-kT}
we have added a new FastJetFinder module within the default ATLAS
card in Delphes. In this FastJetFinder module the parameter $R$ is
set to 1.0 for AK10 jets. Here, we set transverse momentum $p_{T}>200$
GeV for AK10 jets.

\begin{table}
\caption{The cross sections (in pb) of the relevant backgrounds and corresponding
modes with the generated events (where k denotes thousand events units).
\label{tab:The-cross-sections of the background}}

\begin{tabular}{|c|c|c|c|}
\hline 
Background & mode & Cross sections {[}pb{]} & Generated events\tabularnewline
\hline 
\hline 
$t\bar{t}$ & $t\bar{t}$ & $5.990\times10^{2}$ & 179k\tabularnewline
\hline 
\multirow{3}{*}{Single top} & $tj$ & $7.394\times10^{-1}$ & 100k\tabularnewline
\cline{2-4} \cline{3-4} \cline{4-4} 
 & $tb$ & $8.655\times10^{0}$ & 100k\tabularnewline
\cline{2-4} \cline{3-4} \cline{4-4} 
 & $tW$ & $2.553\times10^{-1}$ & 192k\tabularnewline
\hline 
\multirow{3}{*}{Diboson} & $WW$ & $7.937\times10^{1}$ & 100k\tabularnewline
\cline{2-4} \cline{3-4} \cline{4-4} 
 & $WZ$ & $3.021\times10^{1}$ & 100k\tabularnewline
\cline{2-4} \cline{3-4} \cline{4-4} 
 & $ZZ$ & $1.156\times10^{1}$ & 100k\tabularnewline
\hline 
$\begin{array}{c}
W+jets\\
\text{(matched)}
\end{array}$ & $\begin{array}{c}
W+nj\\
n\leq4
\end{array}$ & $2.674\times10^{5}$ & 64k\tabularnewline
\hline 
$\begin{array}{c}
Z+jets\\
\text{(matched)}
\end{array}$ & $\begin{array}{c}
Z+nj\\
n\leq4
\end{array}$ & $8.219\times10^{4}$ & 62k\tabularnewline
\hline 
\end{tabular}
\end{table}

\section{Event Selection}

\label{sec:event-selection-1} 

The analysis searches for top quarks, $W$ bosons, and $b$-jets to
identify vector like VLB/VLX quark candidates that undergo a $B/X\rightarrow Wt$
decay, followed by $t\to Wb$ and $W\rightarrow qq'$ decays. It makes
use of the small-R jets (light jets), large-R jets, and event-based
quantities formed from their combinations. The anti-kt algorithm \citep{Cacciari2008}
implemented in the FastJet package \citep{Cacciari2012} is used to
define two types of jets for this analysis: (1) small-R jets with
$R=0.4$ named AK4 jets, and (2) large-R jets with $R=1.0$ named
AK10 jets. These are reconstructed independently of each other, and
the small-R jets use both tracking information and topological clusters
\citep{ATLAS2017}, while the large-R jets use information from topological
clusters \citep{ATLAS2019} in the calorimeter.

\begin{figure}
\includegraphics[scale=0.4]{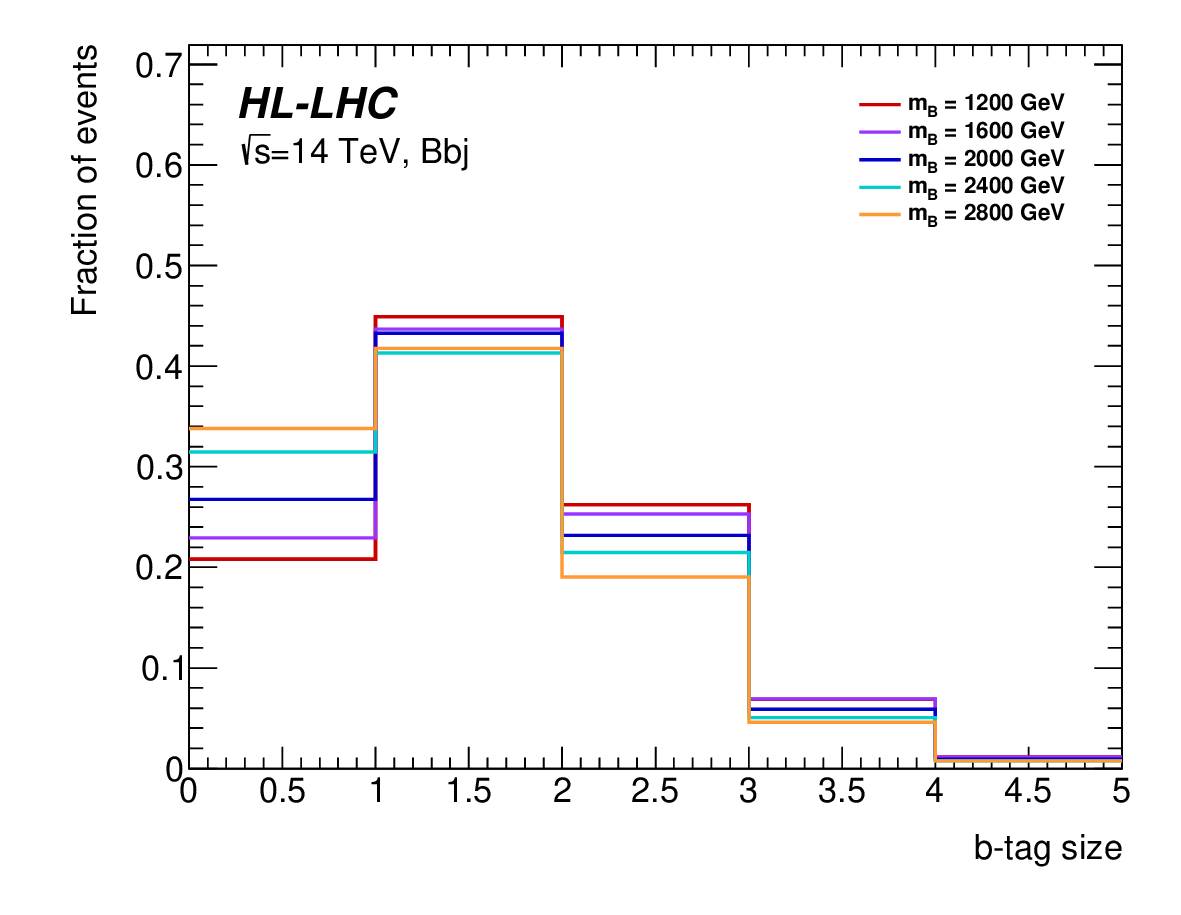}\includegraphics[scale=0.4]{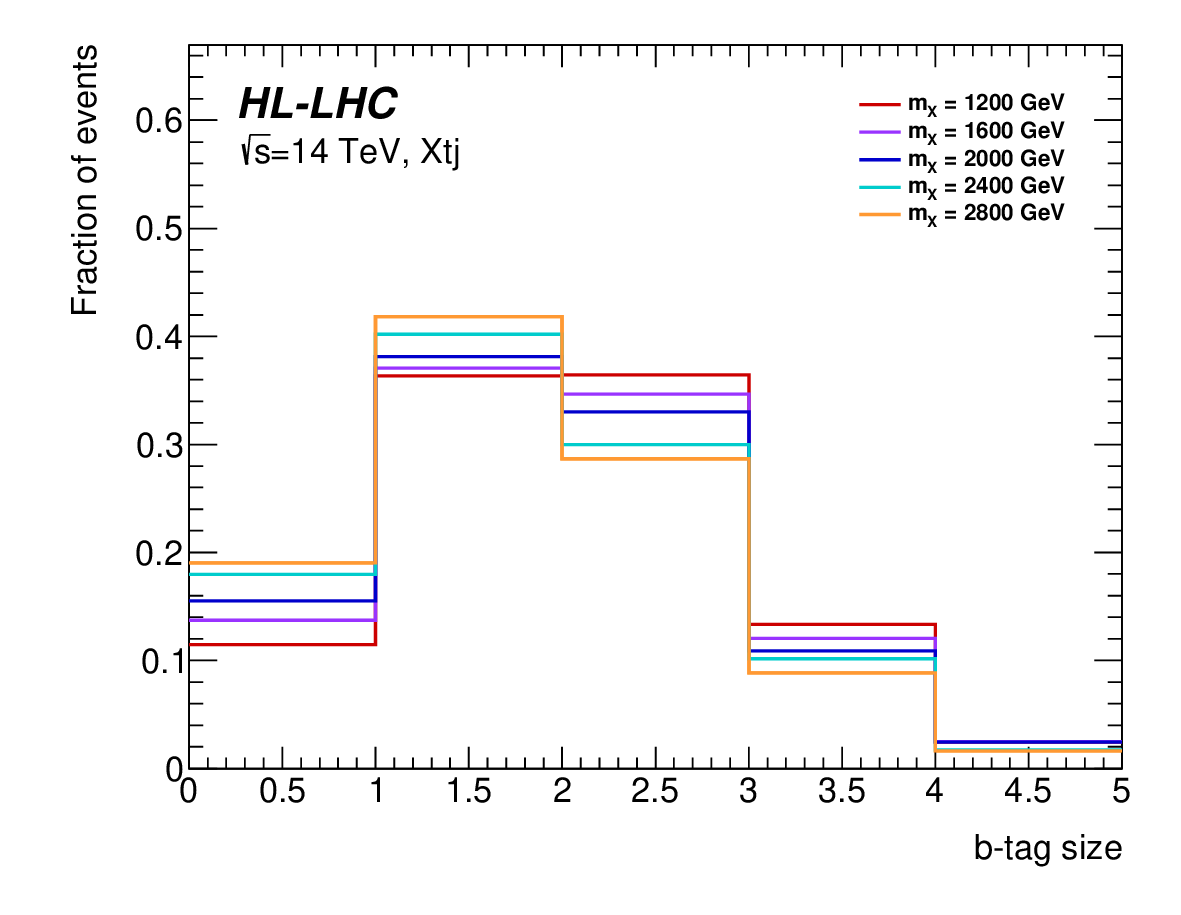}

\includegraphics[scale=0.4]{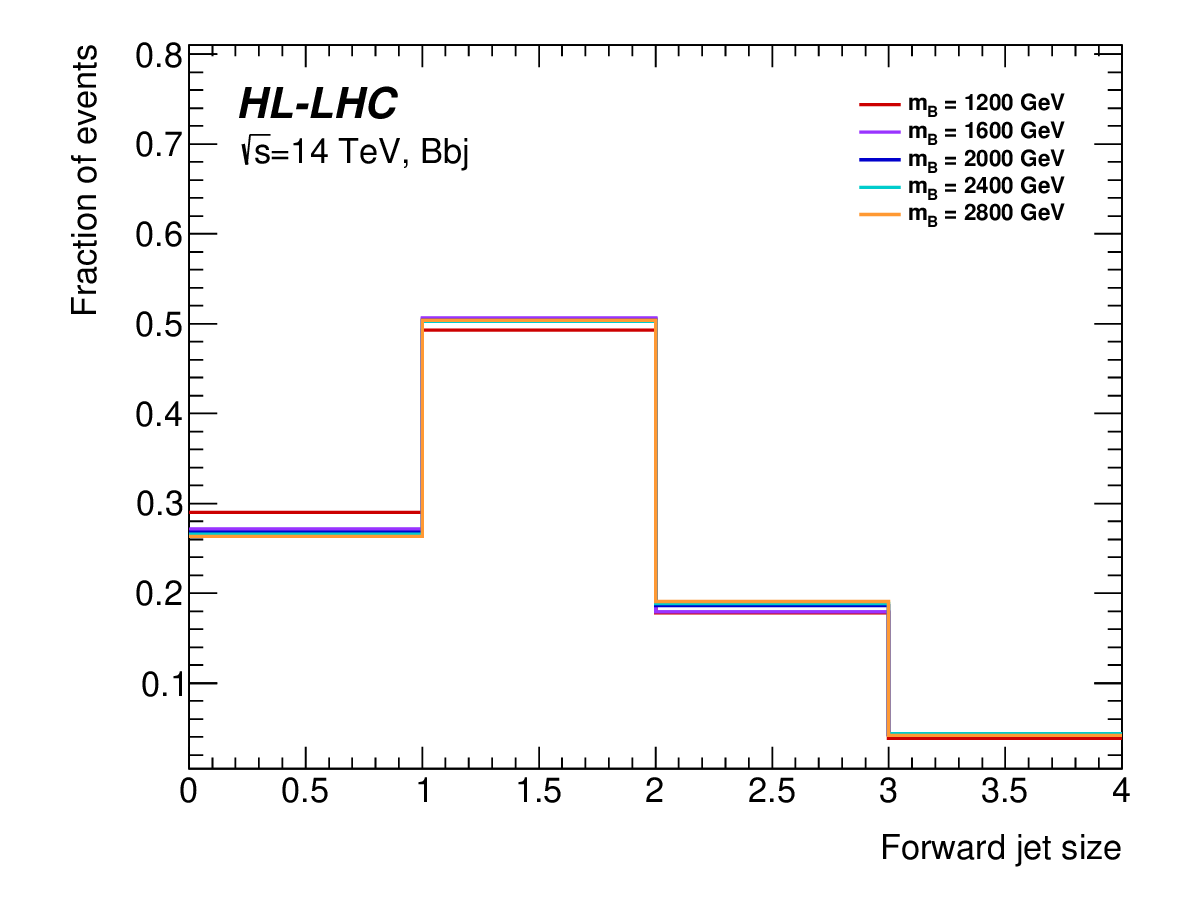}\includegraphics[scale=0.4]{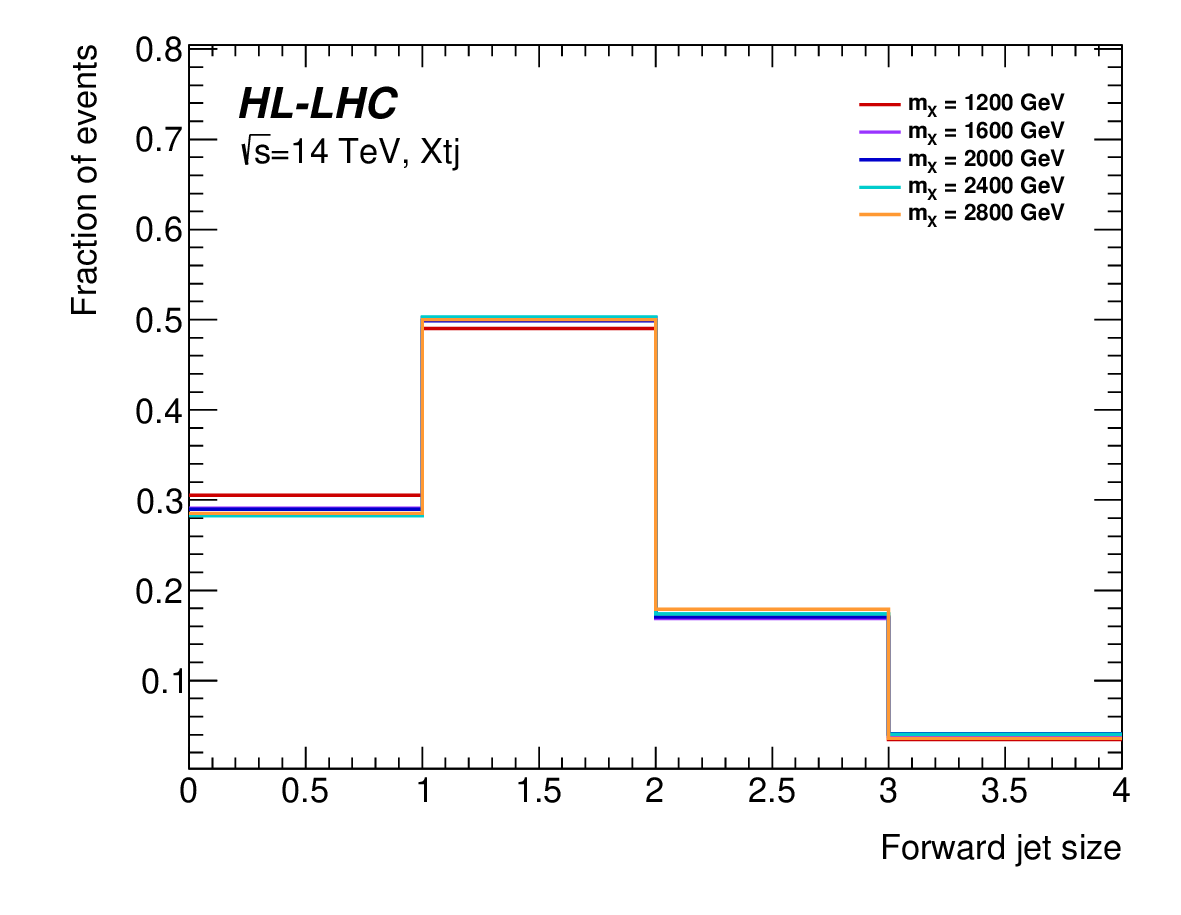}

\caption{The distribution of b-tag size (upper) and forward-jet size (lower)
for different VLB/VLX benchmark mass values. Left (right) panels show
the distributions for $Bbj$ ($Xtj$) production. \label{fig:Figure-caption-2}}
\end{figure}

Jet candidates are required to have $p_{T}>35$ GeV in the forward
region ($2.5<|\eta|<4.5$) and $p_{T}>30$ GeV in the central region
($|\eta|<2.5$). Jets containing $b$-flavored hadrons (“b-jets”)
are used to categorize the events (at least 1 b-tag jet) and reconstruct
the top quark. Simulated distributions of the b-tag jet multiplicity
(size) and the forward-jet size are given in Fig. \ref{fig:Figure-caption-2}
for the benchmark signal processes. Each distribution has been separately
normalized to unity, where the signal samples are from $Bbj$ and
$Xtj$ processes with different VLB/VLX mass values (benchmarks) ranging
from $m_{B}=1200$ GeV to $m_{B}=2800$ GeV.

\begin{figure}
\includegraphics[scale=0.4]{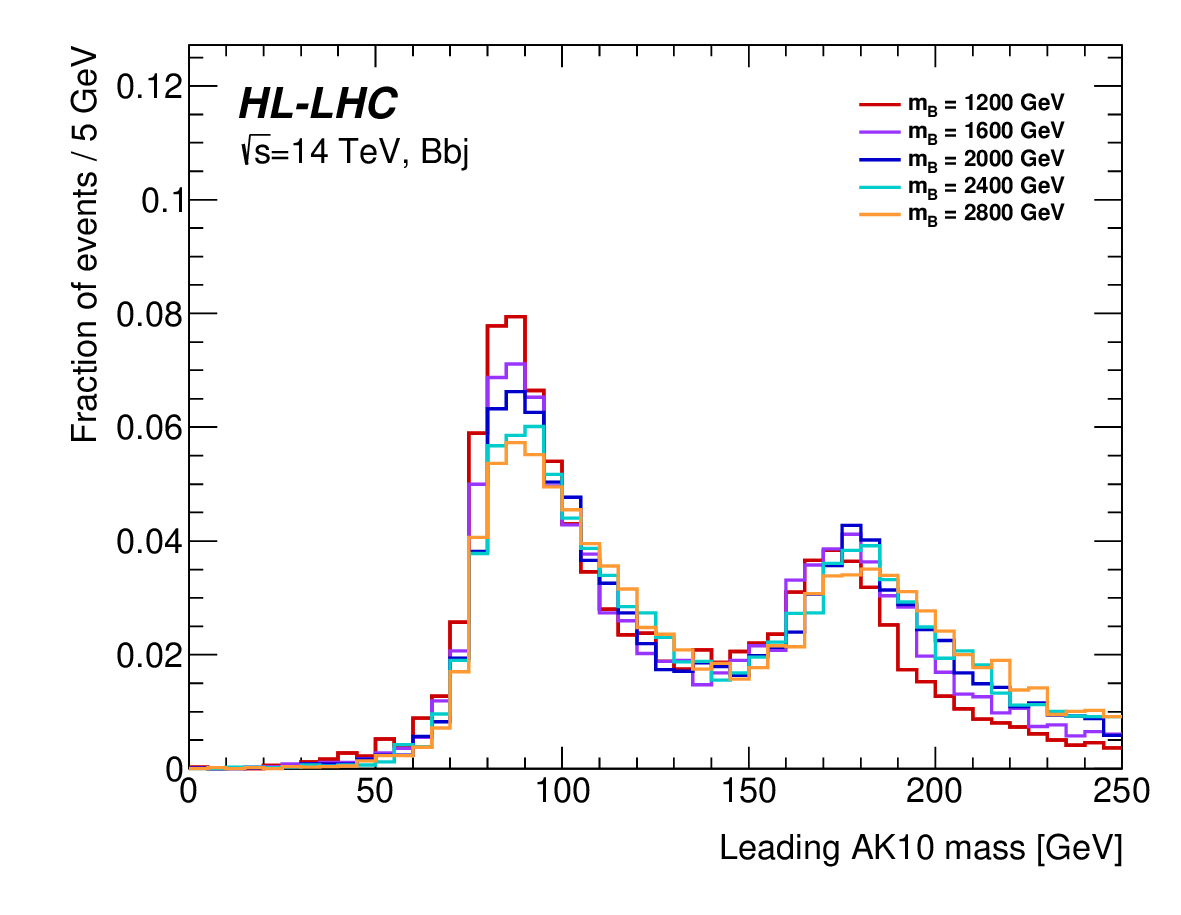}\includegraphics[scale=0.4]{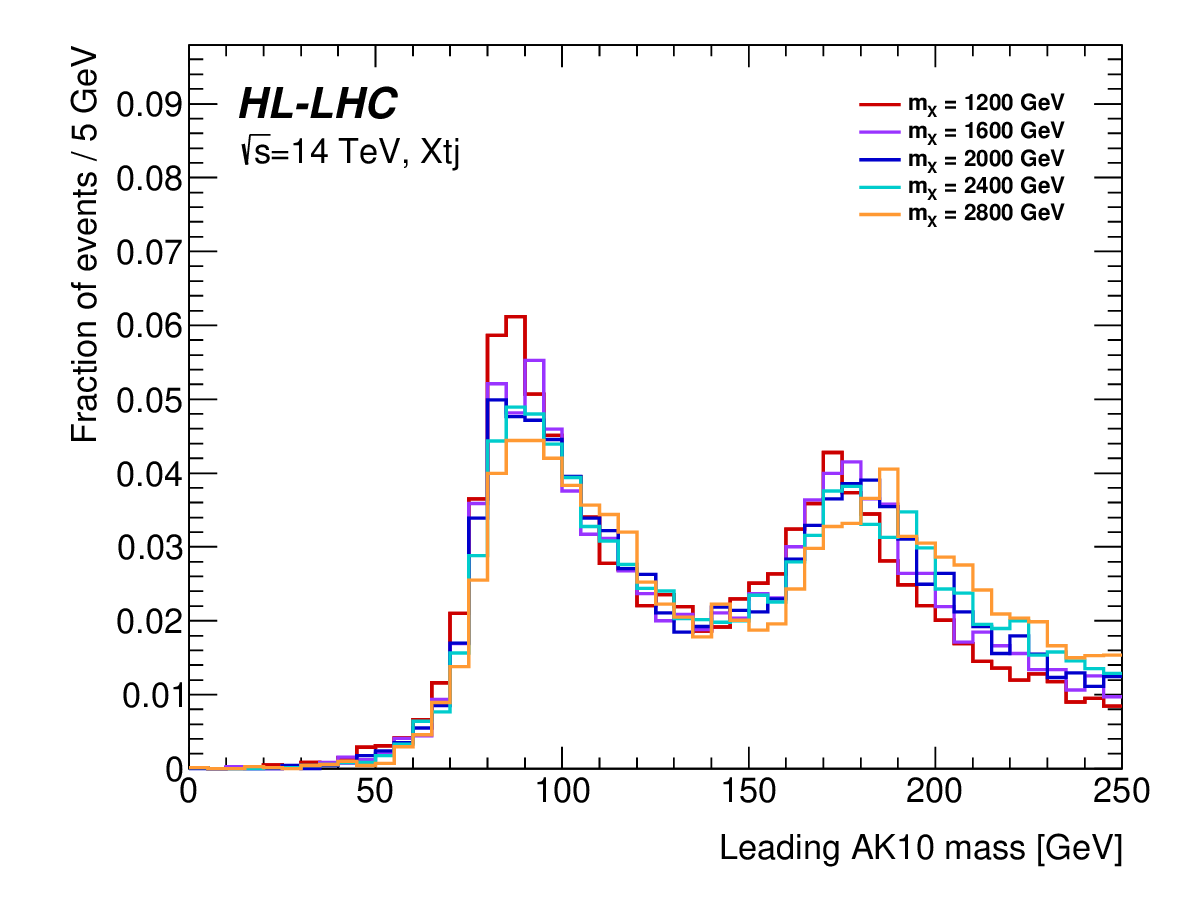}

\includegraphics[scale=0.4]{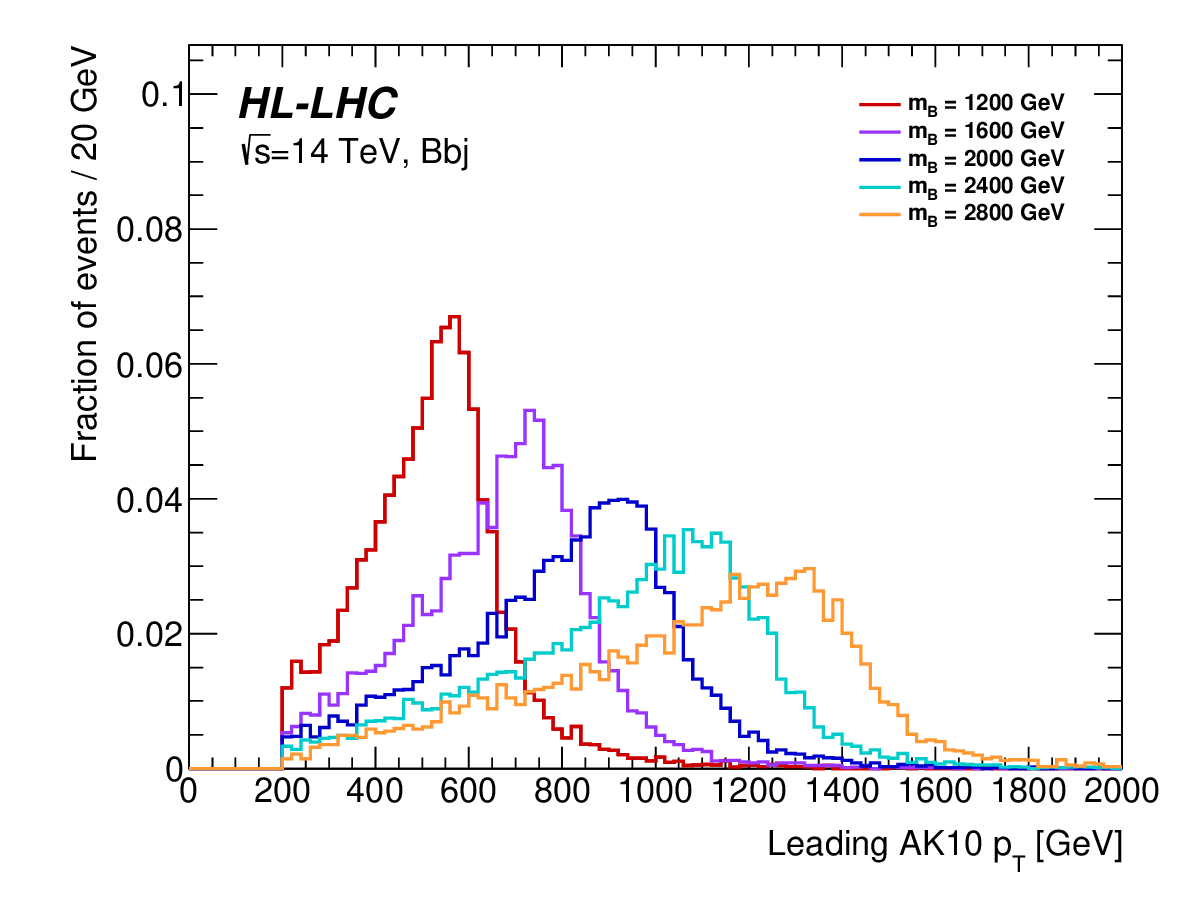}\includegraphics[scale=0.4]{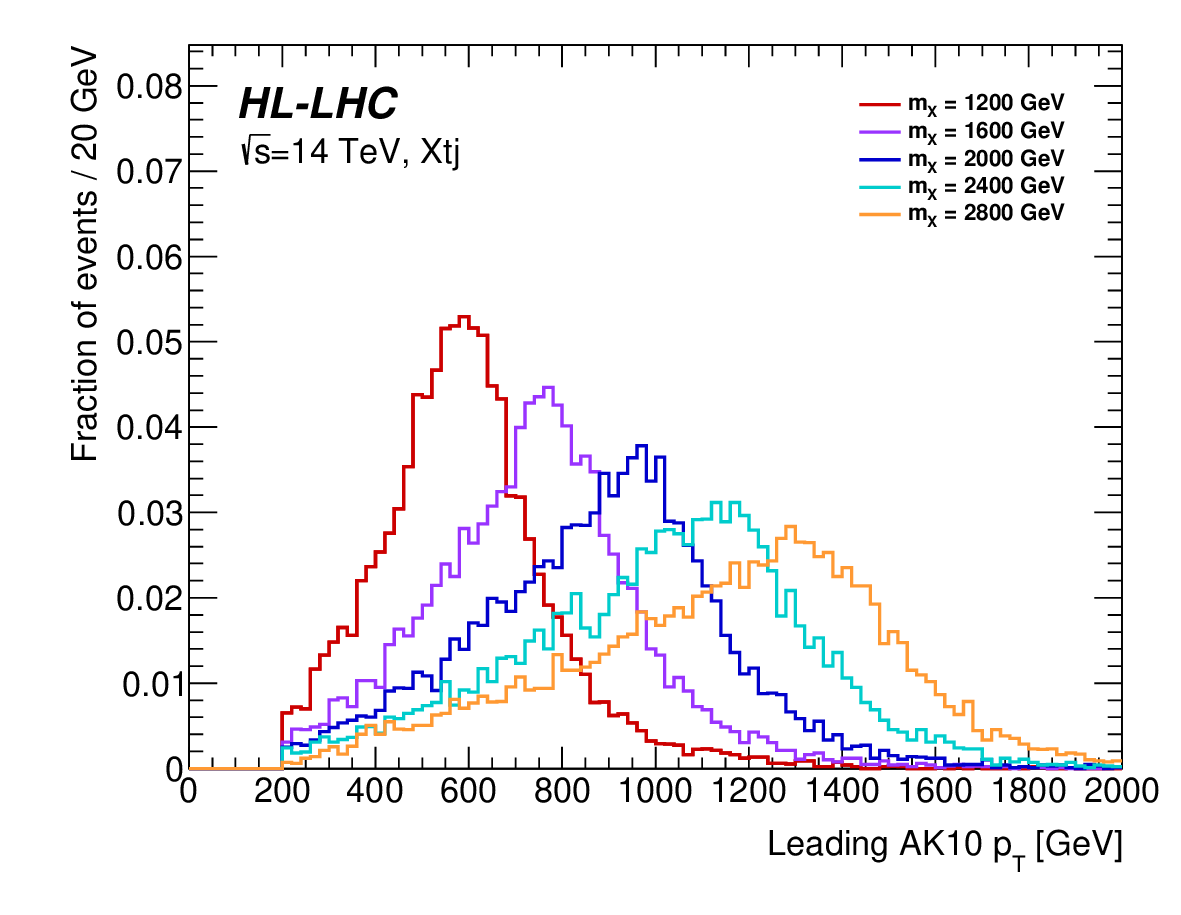}

\caption{The distribution of mass (upper) and transverse momentum $p_{T}$
(lower) of leading AK10 jets. Left (right) panels show the distributions
for $Bbj$ ($Xtj$) production. \label{fig:Figure-caption-3-0}}
\end{figure}

\begin{figure}
\includegraphics[scale=0.4]{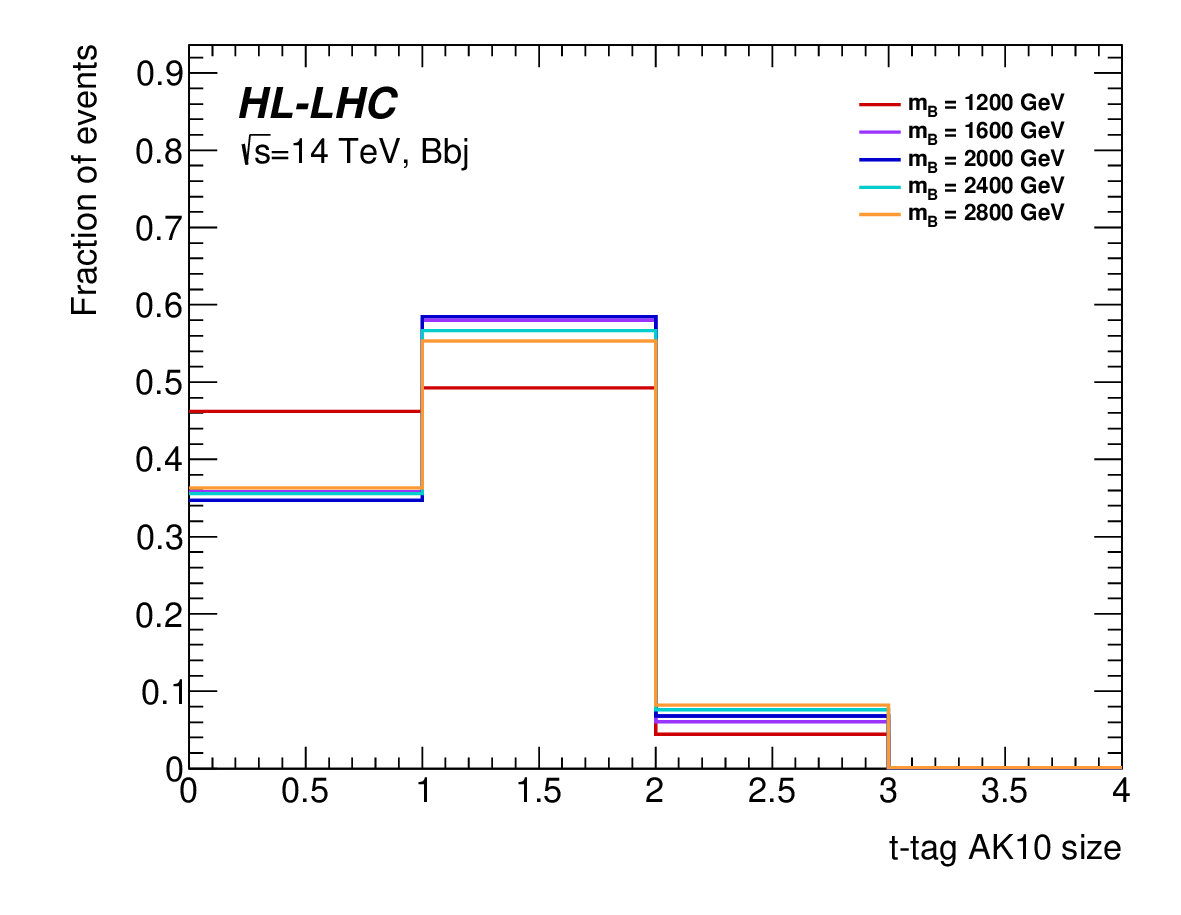}\includegraphics[scale=0.4]{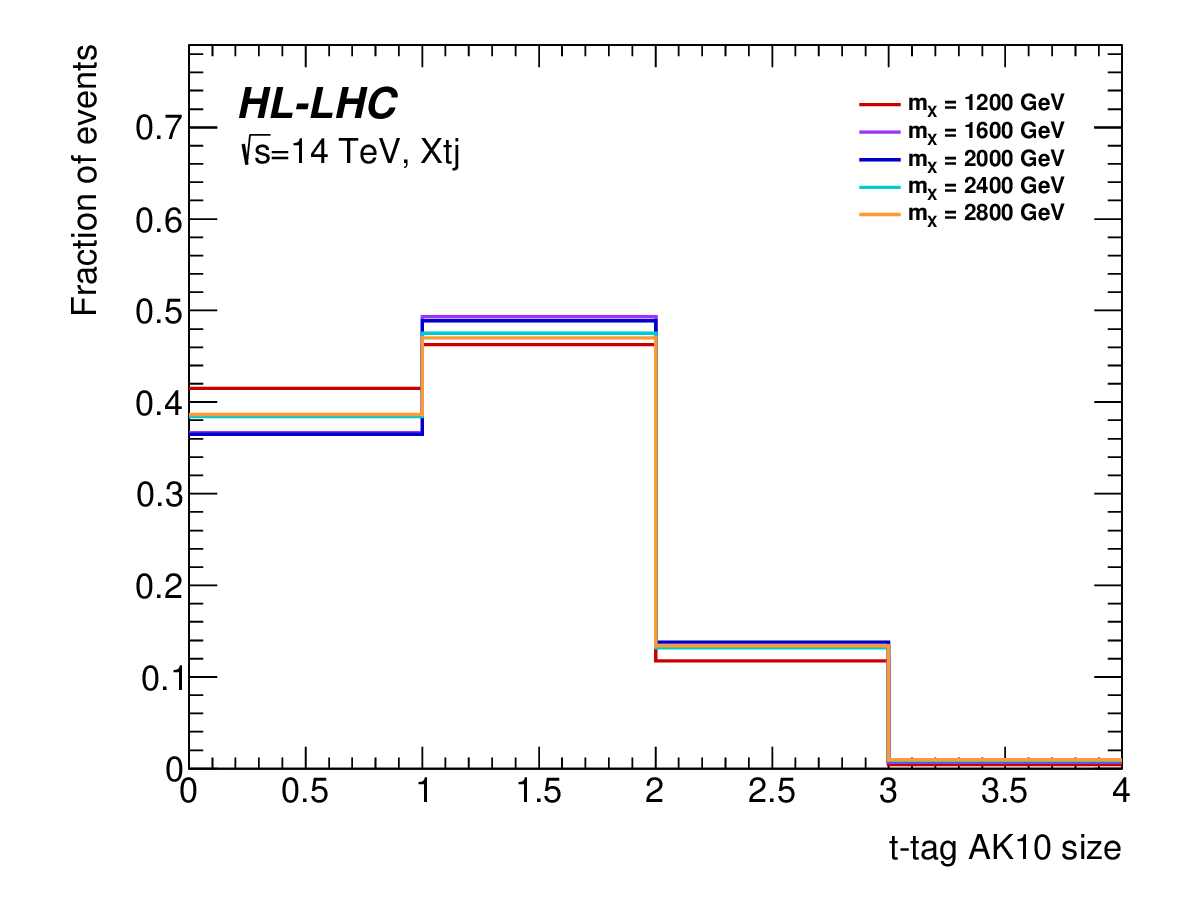}

\includegraphics[scale=0.4]{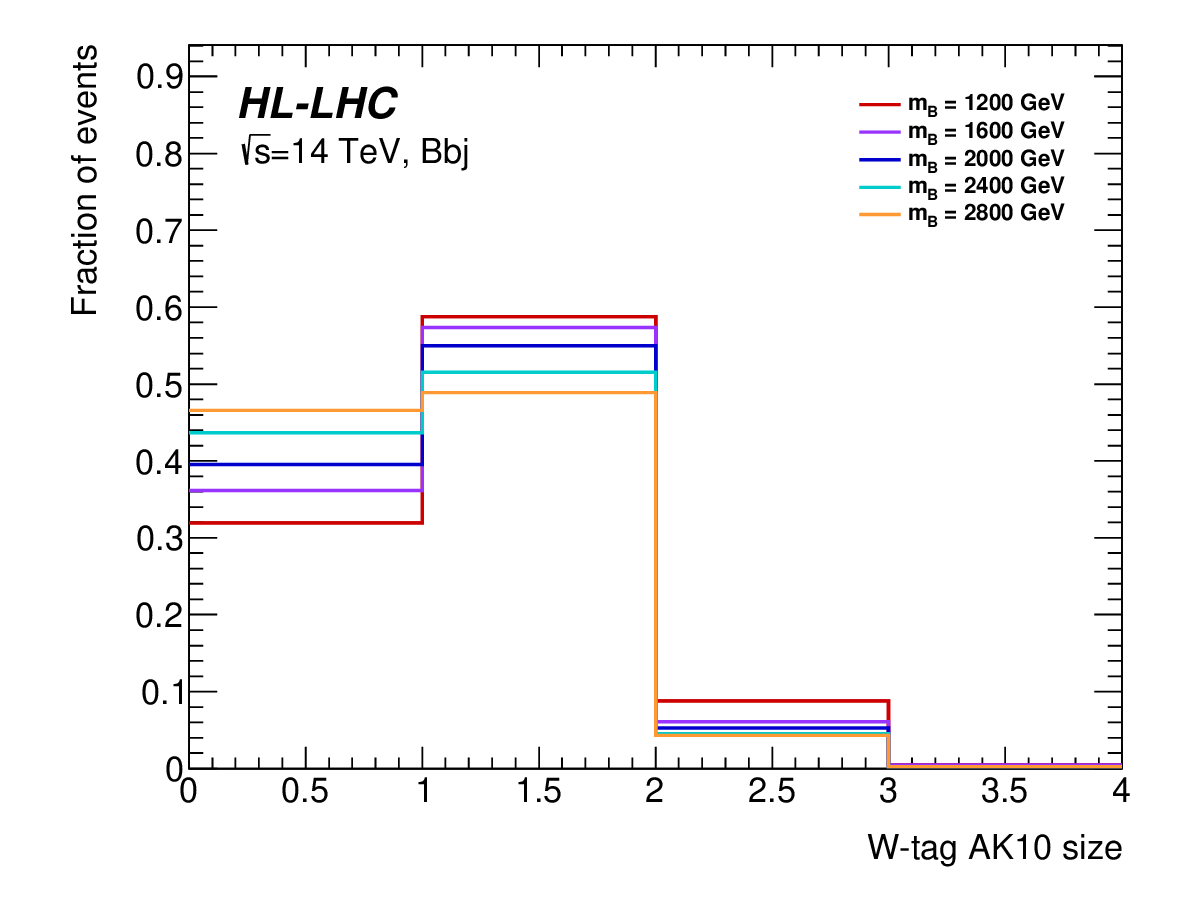}\includegraphics[scale=0.4]{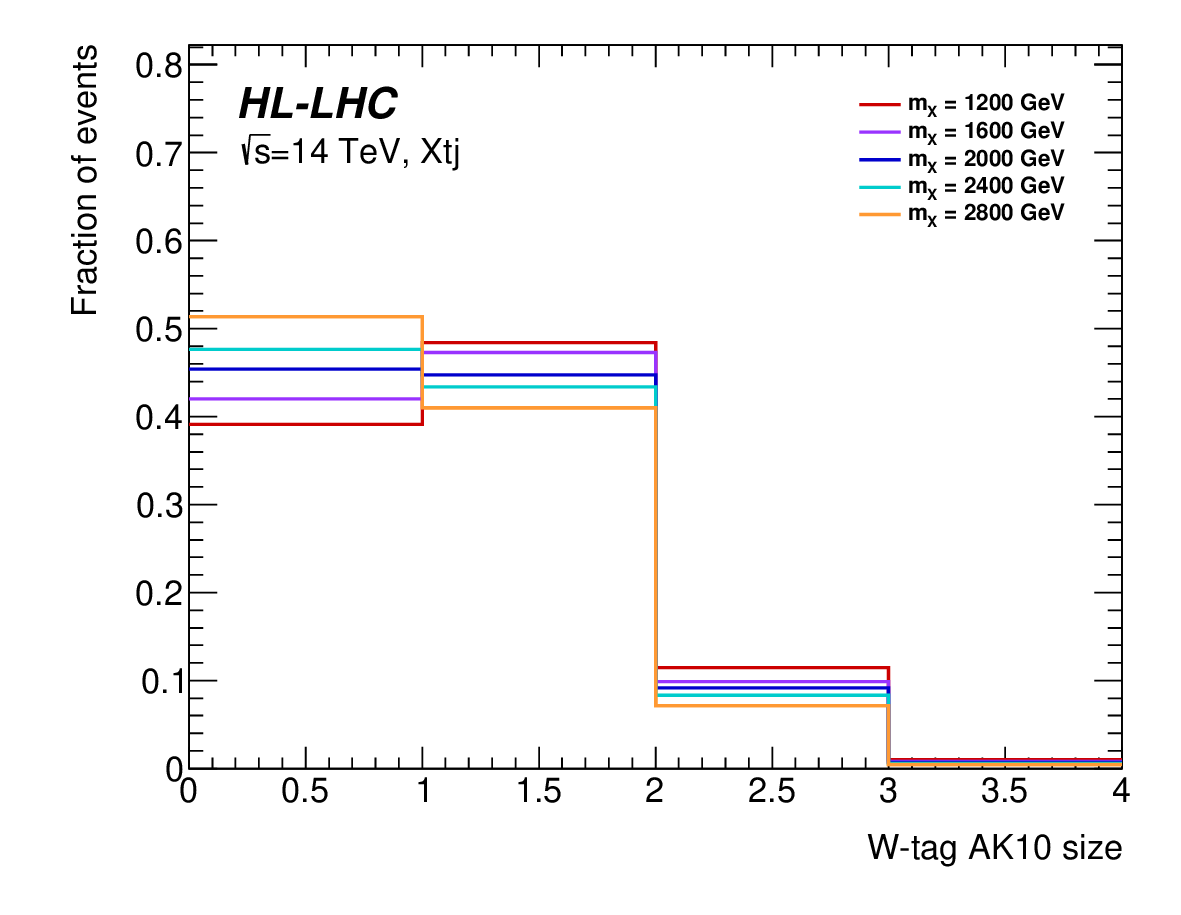}

\caption{The distribution of t-tag size (upper) and W-tag size (lower) for
different VLB/VLX benchmark mass values. Left (right) panels show
the distributions for $Bbj$ ($Xtj$) production. \label{fig:Figure-caption-4}}

\end{figure}

The large-R jet candidates are required to have $|\eta|<2.0$ and
$p_{T}>200$ GeV in addition to their mass interval. The pseudorapidity
$\eta$ cut requirement is imposed to optimize the signal-to-background
ratio and to select jets in a kinematic regime where the object tagging
is efficient and well understood. The distribution of mass and transverse
momentum $p_{T}$ of leading AK10 jets is presented in Fig. \ref{fig:Figure-caption-3-0}.
The $p_{T}$ cut requirement ensures that the large-R jets are selected
efficiently. The events are classified by using tagging states. A
category for the signal are defined including top quark and $W$ boson
tagging state. The identification algorithms (“taggers”) for hadronically
decaying top quarks and $W$ bosons are utilized for the search of
the VLB/VLX quarks in $pp$ collisions at HL-LHC. Distinct tagging
algorithms are employed to identify these different objects. The top
quark tagging states, consisting of quark jets which are clustered
together into the AK10 jets, having $p_{T}>350$ GeV and a mass between
140 and 225 GeV are considered. The $W$ boson candidates are identified
by requiring the AK10 jets, having $p_{T}>200$ GeV and a mass between
60 and 105 GeV. Simulated distributions of the t-tag and W-tag multiplicity
(size) are given in Fig. \ref{fig:Figure-caption-4} for the benchmark
signal processes.

The two highest-$p_{T}$ AK10 jets are referred to as the first leading
and second leading jets. All other AK10 jets are ignored. The event-tagging
states are defined by the three possible tagging states of each AK10
jets as seen in Table \ref{tab:The-event-tagging-states-1}. In the
event selection we take into account one t-tag and one W-tag state.

\begin{table}
\caption{The event-tagging states defined by the tagging states of first leading
and second leading AK10 jets. Relevant major backgrounds are shown
in the table. \label{tab:The-event-tagging-states-1}}

\begin{tabular}{l|c|c|c|c|}
\cline{2-5} \cline{3-5} \cline{4-5} \cline{5-5} 
\multirow{3}{*}{\begin{turn}{90}
$2^{nd}$ AK10 jet
\end{turn}} & \emph{1t 0W} & \textcolor{blue}{Single top} & \textcolor{red}{VLB/VLX} & \textcolor{blue}{Top pair}\tabularnewline
\cline{2-5} \cline{3-5} \cline{4-5} \cline{5-5} 
 & \emph{0t 1W} & \textcolor{blue}{W+jets} & \textcolor{blue}{Dibosons} & \textcolor{red}{VLB/VLX}\tabularnewline
\cline{2-5} \cline{3-5} \cline{4-5} \cline{5-5} 
 & \emph{0t 0W} & no-tag & \textcolor{blue}{W+jets} & \textcolor{blue}{Single top}\tabularnewline
\cline{2-5} \cline{3-5} \cline{4-5} \cline{5-5} 
\multicolumn{1}{l}{} &  & \emph{0t 0W} & \emph{0t 1W} & \emph{1t 0W}\tabularnewline
\cline{3-5} \cline{4-5} \cline{5-5} 
\multicolumn{1}{l}{} & \multicolumn{1}{c}{} & \multicolumn{3}{c}{$1^{st}$ AK10 jet}\tabularnewline
\end{tabular}
\end{table}

With these tagging definitions, the events are classified according
to the tagging states of each AK10 jet: the large-R jet could be top-quark
tagged, be W-boson tagged or be neither W-boson tagged nor top-quark
tagged. In our case, a signal event can be in two entries in a $3\times3$
matrix defined in Table \ref{tab:The-event-tagging-states-1} to categorize
possible tagging states of the two ($1^{st}$ and $2^{nd}$) AK10
jets in an event. The signal region (VLB/VLX) consists of $tW$-tag
state as illustrated with red colors in Table \ref{tab:The-event-tagging-states-1}.
However, $2t$-tag state or $2W$-tag state is not considered in the
signal region since we search single VLB/VLX in $Wt$ decay channel,
where one of the hadronic W boson and b-tag quark taken into account
as the constituents of the top tagged events. Moreover, one of the
main characteristics of the signal is the forward jet, which is benefited
to separate signal from background.

For a correct reconstruction of VLB/VLX signal, we have applied topological
cuts on angular separation $\Delta\phi$ and angular distance $\Delta R=\sqrt{(\Delta\eta)^{2}+(\Delta\phi)^{2}}$.
Here, we take the cut $|\Delta\phi(t,W)|>2$ for verifying balance
between the transverse momentum of t-tag jet and W-tag jet. However,
the cuts $\Delta R(t,b)<1$ and $\Delta R(W,b)>2$ are used to verify
that b-tag jet comes from top decay. 

The reconstruction of events with a $t$-tag and $W$-tag is found
to be best suited for each benchmark parameter values (within high
mass range) of the VLQs. The $\chi^{2}$ method provide a stable performance
for all VLQ masses, where the VLB/VLX is reconstructed from the t-tag
jet and $W$-tag jet. 

\[
\chi^{2}=\frac{(\Delta R(t,W)-\pi)^{2}}{\sigma_{\Delta R}^{2}}+\frac{\left[(p_{T}^{t}-p_{T}^{W})/(p_{T}^{t}+p_{T}^{W})\right]^{2}}{\sigma_{p_{T}}^{2}}
\]
For the $\chi^{2}$ quantity the angular distance $\Delta R(t,W)$
and the $p_{T}$ balance are used. It is verified in simulation that
the expected values of $\Delta R(t,W)$ and the $p_{T}$ balance are
$\pi$ and 0, with their standard deviations $\sigma_{\Delta R}$
and $\sigma_{pT}$, respectively. 

The main backgrounds from top pair and single top productions, W/Z+jets
and Dibosons contributing backgrounds are evaluated at these categorized
signal region. Together with the top quark tagged jets and W-boson
tagged jets, the jets with b-tag are also included, as these typically
result from mistagging a charm-quark jet arising from a CKM favored
decay $W^{+}\to c\bar{s}$ (or $W^{-}\to\bar{c}s$). 

A summary of event selection criteria used in the analysis has been
shown in Table \ref{tab:Cuts table}.

\begin{table}

\caption{Summary of event selection criteria applied in the analysis. \label{tab:Cuts table}}

\begin{tabular}{|c|cc|c|c|}
\hline 
Selection type & Value &  & Selection type & Value\tabularnewline
\hline 
\hline 
Electron size & $=0$ &  & AK10 jet size & $\geq2$\tabularnewline
\hline 
Muon size & $=0$ &  & AK10 jet $|\eta|$ & $<2$\tabularnewline
\hline 
AK4 jet size & $\geq2$ &  & t-tagged jet $p_{T}$ & $>350$ {[}GeV{]}\tabularnewline
\hline 
Forward jet size & $\geq1$ &  & t-tagged jet mass interval & $140<m_{AK10}<225$ {[}GeV{]}\tabularnewline
\hline 
Forward jet $p_{T}$ & $>35$ {[}GeV{]} &  & W-tagged jet $p_{T}$ & $>200$ {[}GeV{]}\tabularnewline
\hline 
Forward jet $\eta$ & $>2.5$ &  & W-tagged jet mass interval & $60<m_{AK10}<105$ {[}GeV{]}\tabularnewline
\hline 
b-tag size & $=1$ &  & $\Delta\phi(t,W)$ & $>2$\tabularnewline
\hline 
b-tag $p_{T}$ & $>30$ {[}GeV{]}  &  & $\Delta R(t,b)$ & $<1$\tabularnewline
\hline 
b-tag $|\eta|$ & $<2.4$ &  & $\Delta R(W,b)$ & $>2$\tabularnewline
\hline 
\end{tabular}

\end{table}

\section{Reconstruction and Statistics}

\label{sec:reconstruction and statistics-1} 

We assume the existence of single VLB/VLX quark production in different
production modes ($Bbj$ and $Btj/Xtj$), the signal would appear
as an excess of events with $Wt$ invariant masses around the VLB/VLX
quark mass. Reconstructed invariant mass ($m_{B/X}$) of VLB/VLX quarks
is defined as $m_{B/X}=\sqrt{(p_{t}+p_{W})^{2}}=\sqrt{(m_{t}^{2}+m_{W}^{2}+2|\mathbf{p}_{t}|\cdot|\mathbf{p}_{W}|\cos\theta}$.
Here, the $p_{t}$ and $p_{W}$ are four-momentum, the $\mathbf{p}_{t}$
and $\mathbf{p}_{W}$ are momentum vectors of top quark and W-boson,
respectively. The most probable value for the $\cos\theta$ can be
set, since these two objects are mostly back to back, because of the
heavy object resonant production. However, in the analysis of signal
and background samples we use the physics four-vectors within the
Root 6 \citep{Root6}, and the analysis code has been developed using
Python interface PyRoot which is able to interoperate with widely-used
Python data-science libraries.

\begin{figure}
\includegraphics[scale=0.4]{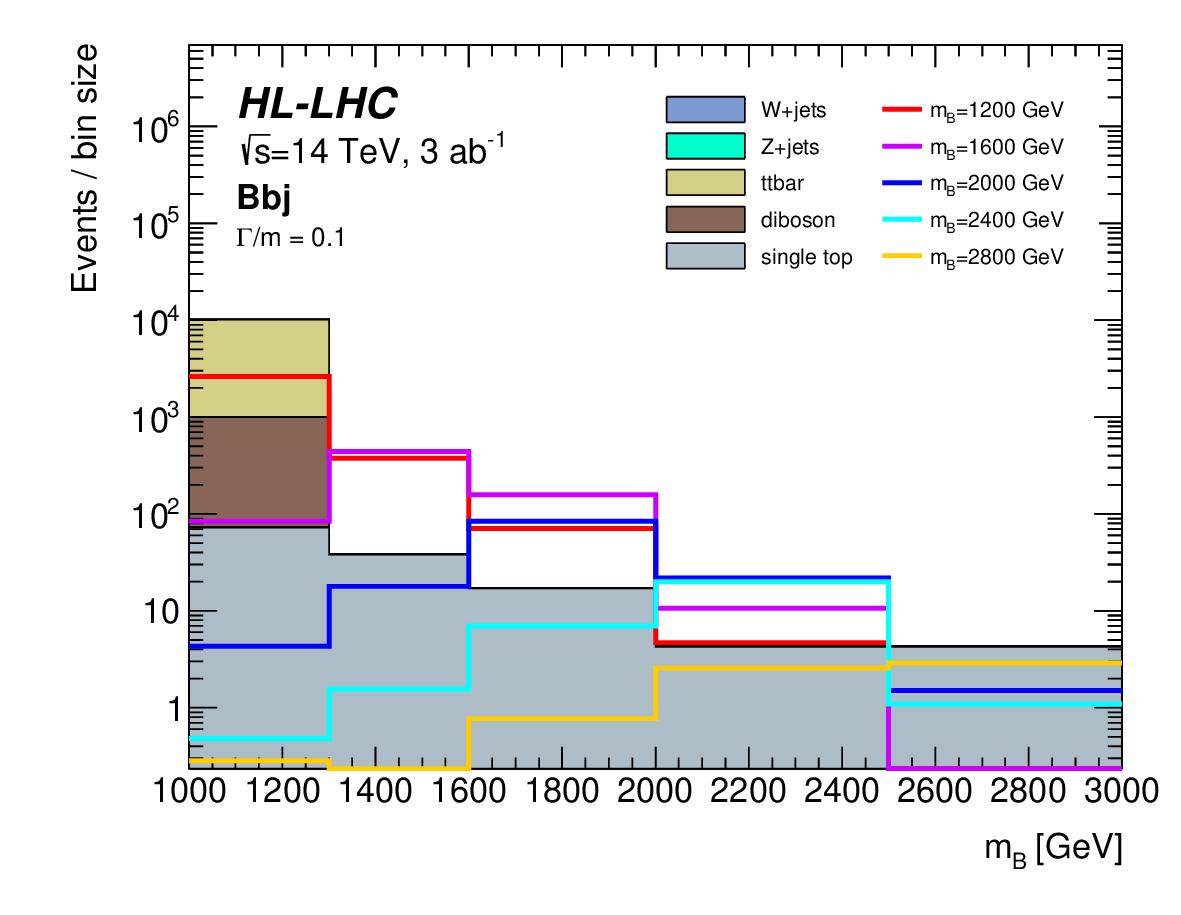}\includegraphics[scale=0.4]{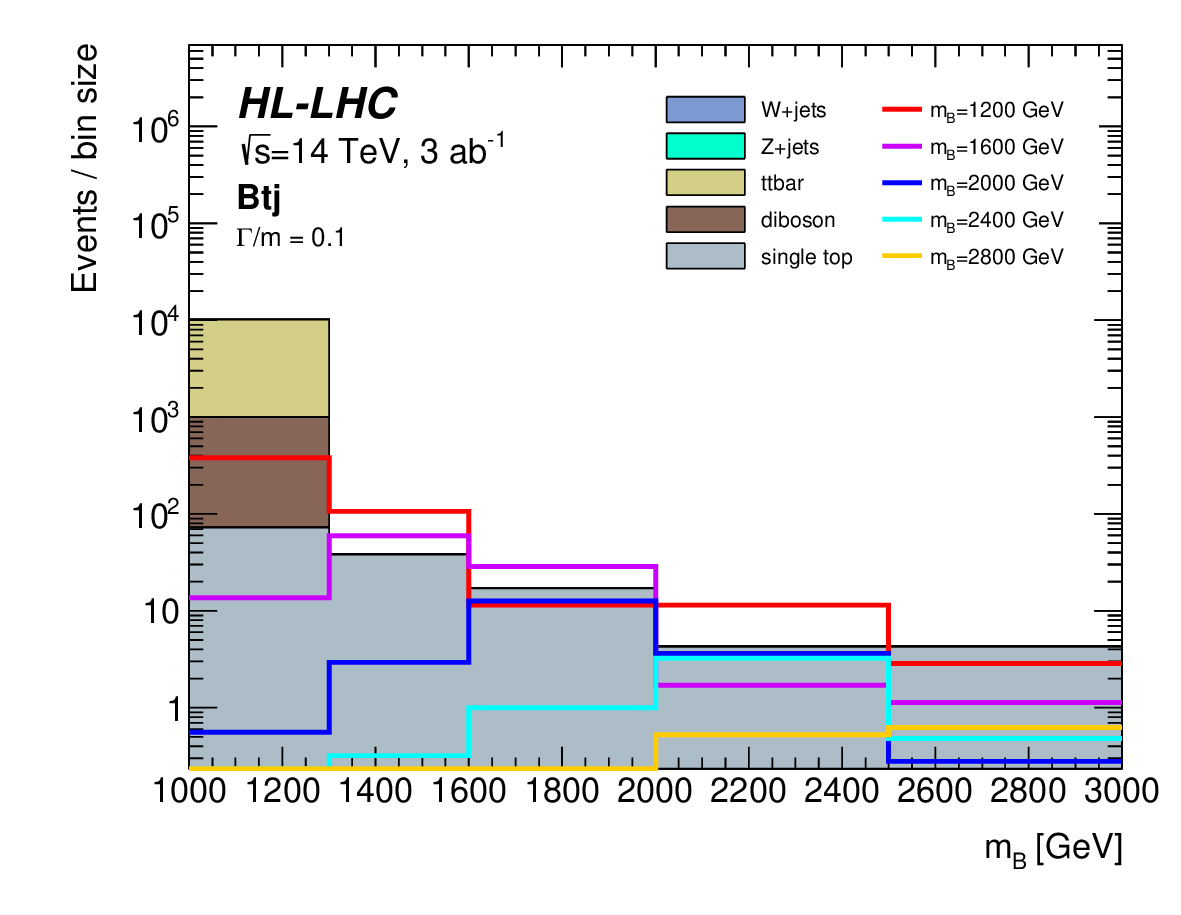}

\includegraphics[scale=0.4]{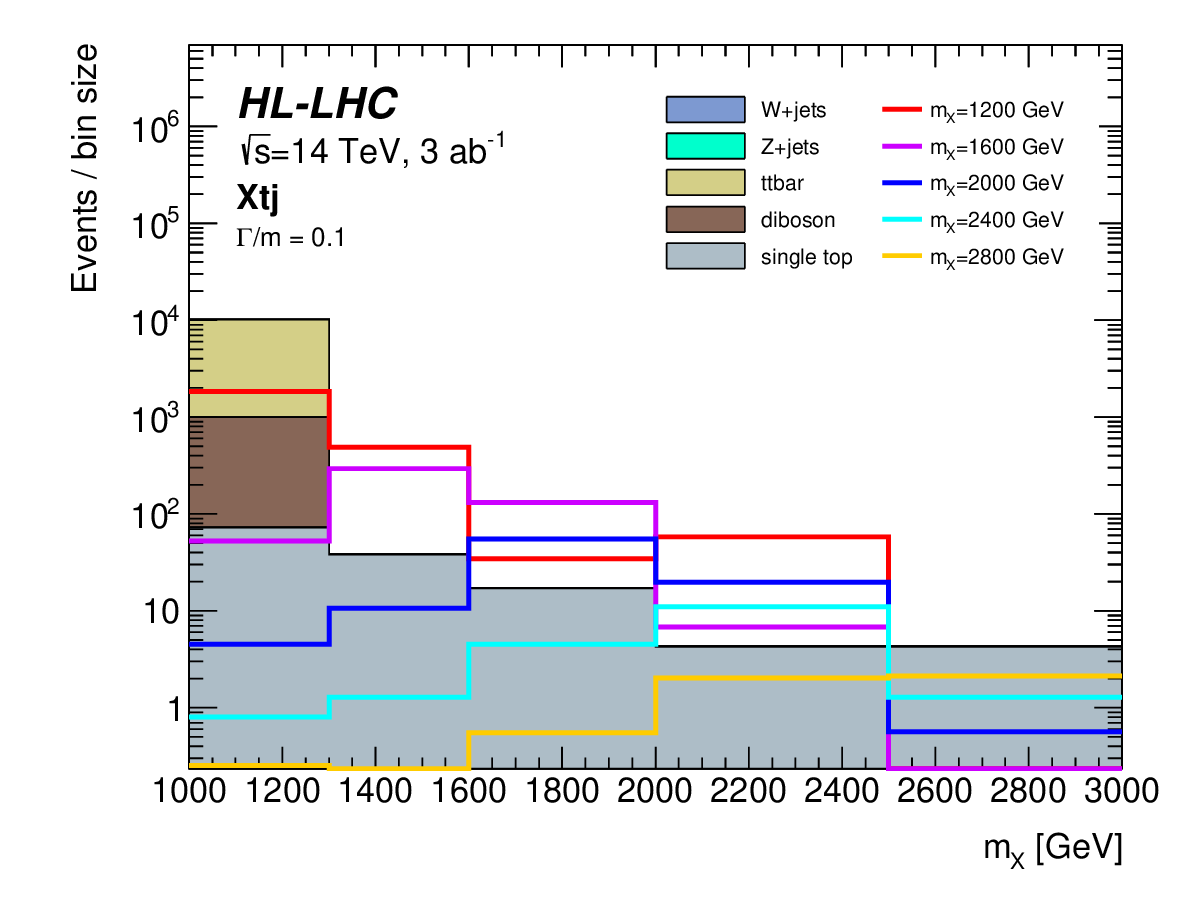}

\caption{Invariant mass ($m_{B/X}$) of $tW$ system reconstructing VLB (upper:
Bbj and Btj production) and VLX (lower: Xtj production) signal. The
corresponding SM backgrounds are shown in stack plot. \label{fig:Reconstructed-mass-of}}
\end{figure}

The reconstructed mass distributions for the signal category is presented
in Fig. \ref{fig:Reconstructed-mass-of}. In the figure, the left
panel shows $Bbj$ production mode, while right panel shows the production
mode $Xtj$. From the invariant mass distributions for $Bbj$ and
$Btj/Xtj$ processes the number of expected events for signal and
background have been calculated in the mass window $10\%$ of the
VLB/VLX mass values.

In order to analyze the sensitivity, we use the statistical significance
($SS$) for expected discovery ($SS_{dis}$) limits \citep{Glen2011}

\[
SS_{dis}=\sqrt{2\left[(S+B)\ln\left(\frac{(S+B)(1+\delta_{sys}^{2}B)}{B+(S+B)\delta_{sys}^{2}B}\right)-\frac{1}{\delta_{sys}^{2}}\ln\left(1+\frac{\delta_{sys}^{2}S}{1+\delta_{sys}^{2}B}\right)\right]}
\]

and exclusion ($SS_{exc}$) limits

\[
SS_{exc}=\sqrt{2\left[S-B\ln\left(\frac{(S+B+X)}{2B}\right)-\frac{1}{\delta_{sys}^{2}}\ln\left(\frac{B-S+X}{2B}\right)\right]-(B+S-X)\left(1+\frac{1}{\delta_{sys}^{2}B}\right)}
\]

with 

\[
X=\sqrt{(S+B)^{2}-4S\delta_{sys}^{2}B^{2}/(1+\delta_{sys}^{2}B)}
\]

Here, $S$ and $B$ are the expected number of events for the signal
and background, respectively. These can be obtained by multiplying
the production cross sections, branchings and the integrated luminosity
together with the corresponding efficiencies for interested search
channel. The systematic uncertainty ($\delta_{sys}$ in percentage)
on the estimated SM background is already included in the significance.
However, in the limit case ($\delta_{sys}\to0$) we obtain these expressions
as $SS_{dis}=\sqrt{2[(S+B)\ln(1+S/B)-S]}$ and $SS_{exc}=\sqrt{2[S-B\ln(1+S/B)]}$
as already used in many of the phenomenological studies.

\begin{figure}
\includegraphics[scale=0.4]{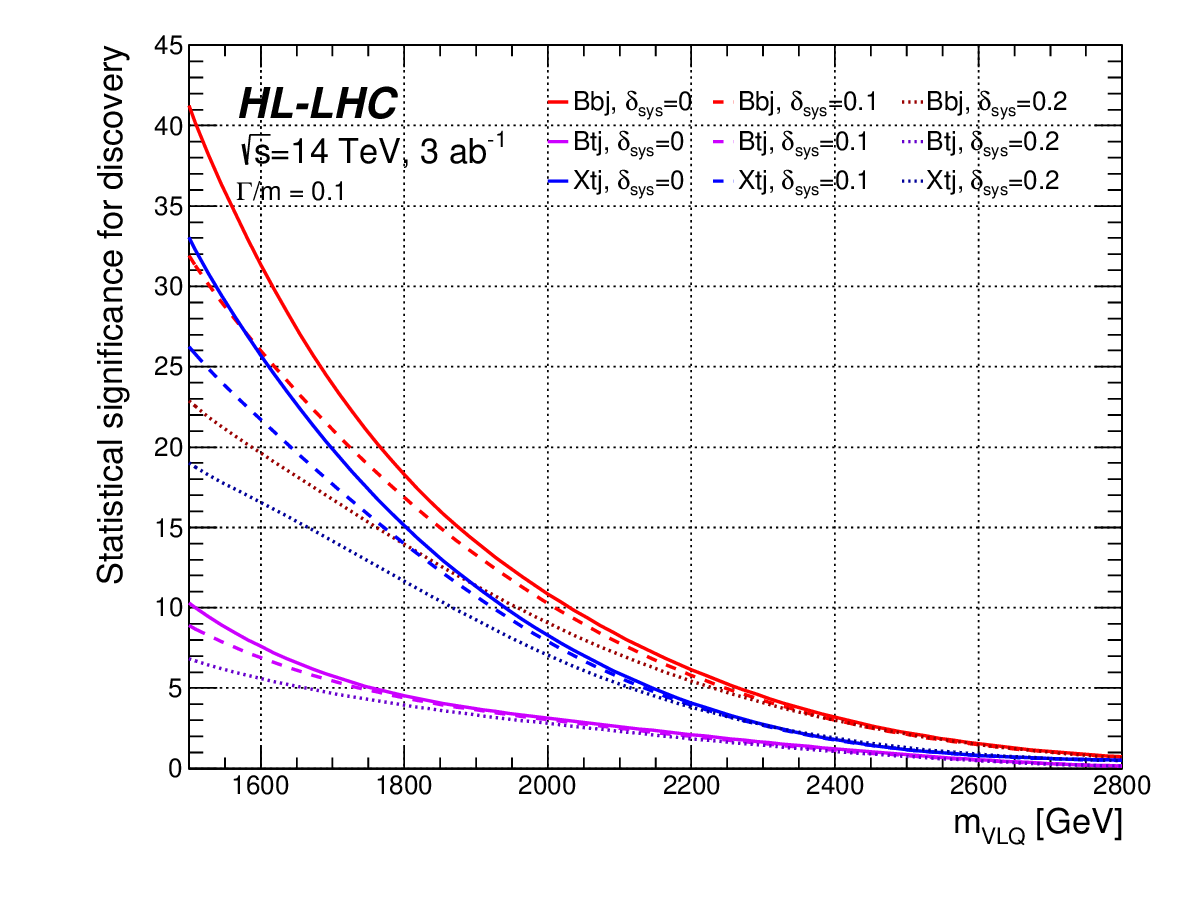}\includegraphics[scale=0.4]{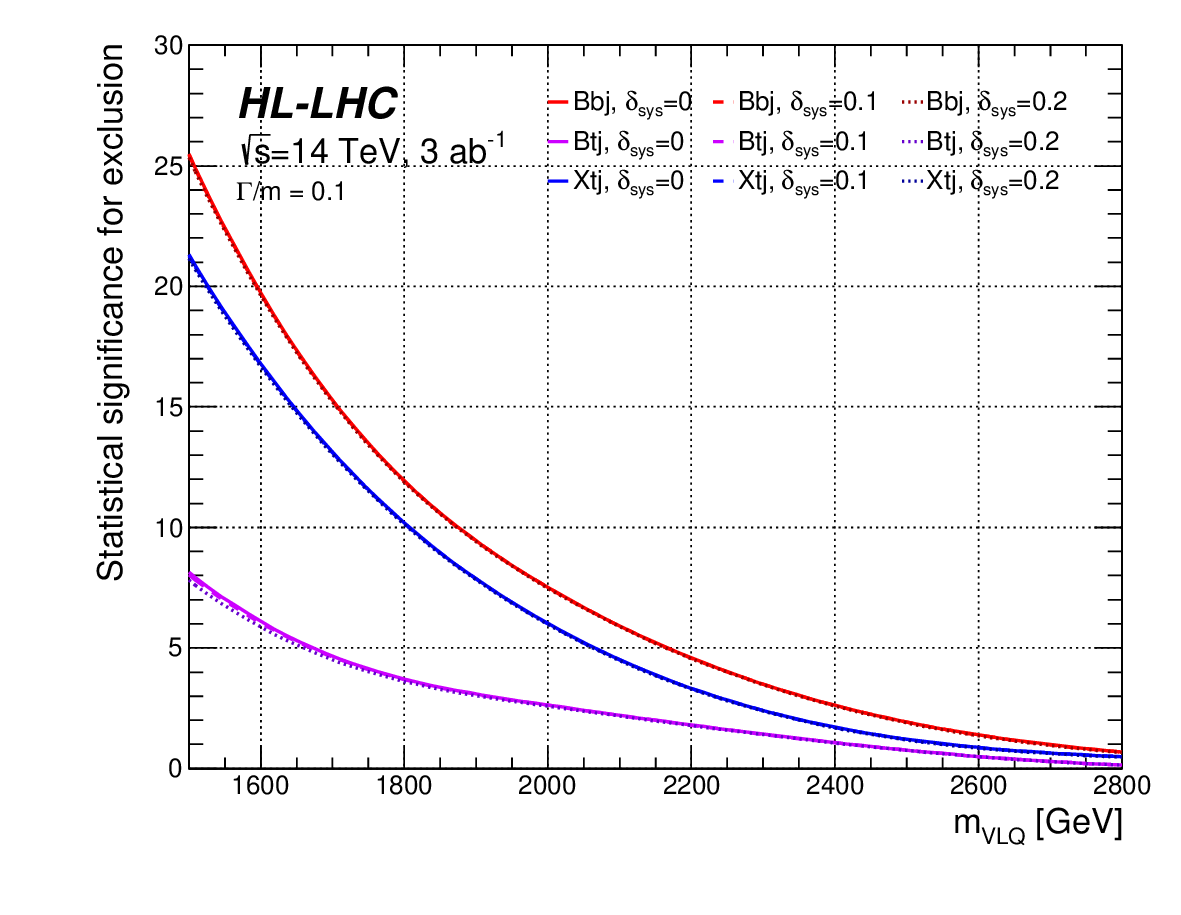}

\includegraphics[scale=0.4]{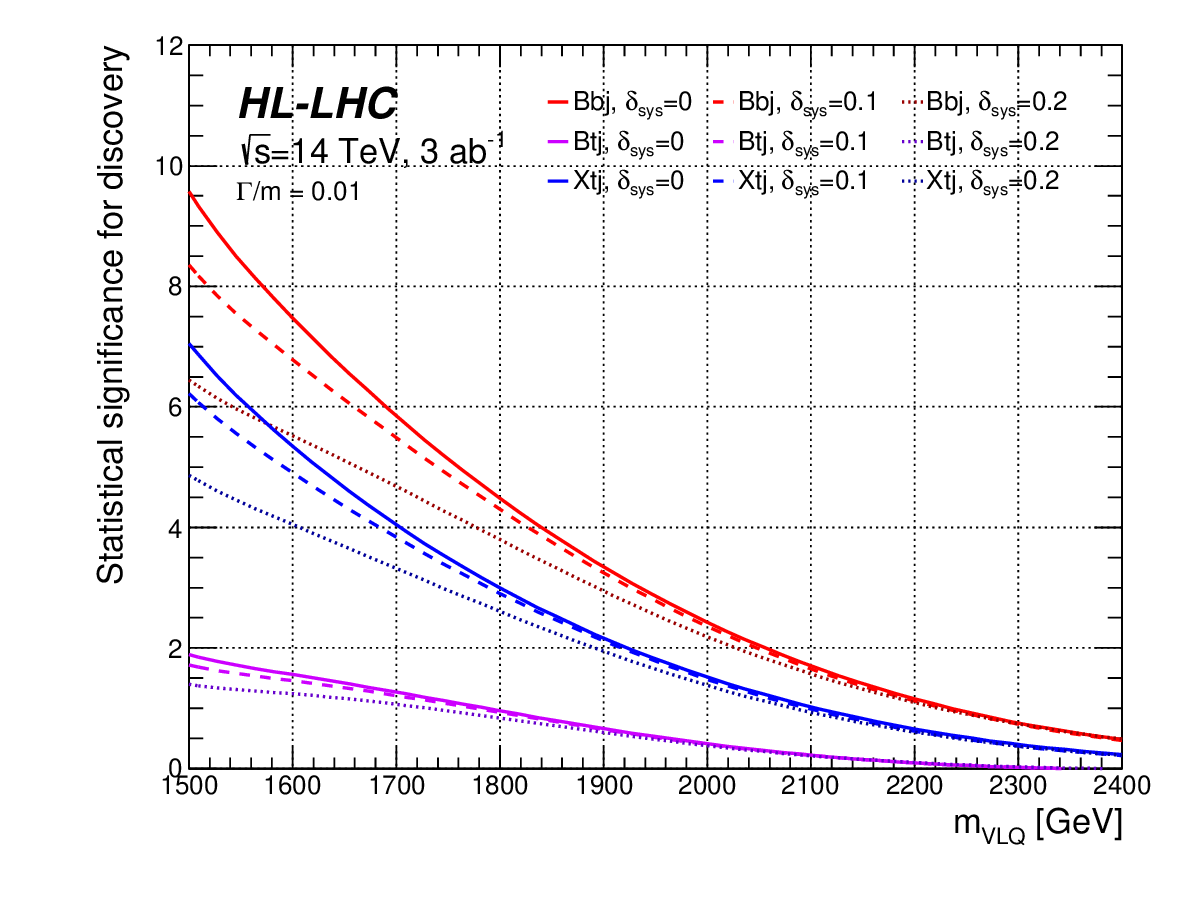}\includegraphics[scale=0.4]{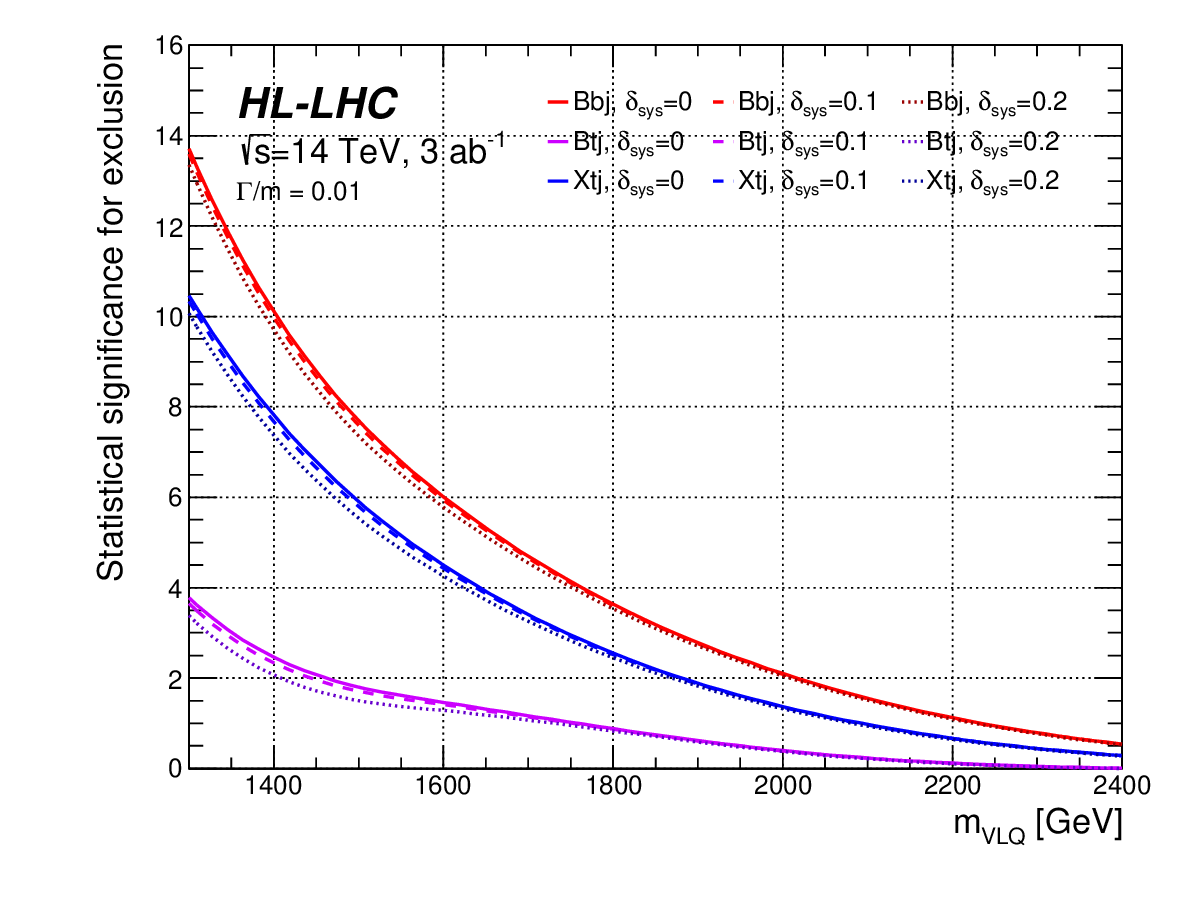}

\caption{Significance for the discovery (left) and exclusion (right) of VLB/VLX
signal for the production process $Bbj$ and $Btj/Xtj$. \label{fig:Significance-for-the}}

\end{figure}

As a result, in Fig. \ref{fig:Significance-for-the}, we present exclusion
plot and discovery plot with three different systematic uncertainty
cases: no systematics ($\delta_{sys}\to0$), a mild systematic of
$\delta_{sys}=10\%$, and an possible systematic of $\delta_{sys}=20\%$.
The mass limits from the discovery plot is effected by the systematic
uncertainties. One can see that with a possible uncertainty of $20\%$,
sensitivities are slightly weaker than those with a mild systematic
uncertainty of $10\%$. For a width to mass ratio of $\Gamma_{B/X}/M_{B/X}=0.1$,
the VLB (VLX) can be discovered (with $5\sigma$ level) with a mass
about 2273 (2145) GeV at HL-LHC with an integrated luminosity of 3
ab$^{-1}$. Out of a discovery, one can also set $95\%$ CL exclusion
limits 2491 (2364) GeV on the mass of VLB (VLX) for the same integrated
luminosity. For more narrow width case as $\Gamma_{B/X}/M_{B/X}=0.01$,
the VLB (VLX) can be discovered (with $5\sigma$ level) with a mass
about 1764 (1618) GeV at HL-LHC with an integrated luminosity of 3
ab$^{-1}$. Out of a discovery, one can also set $95\%$ CL exclusion
limits 2018 (1873) GeV on the mass of VLB (VLX) for the same integrated
luminosity. Mass limits on VLB/VLX for three different production
processes and systematic uncertainty cases are shown in Table \ref{tab:Limit-table}.
For a comparison with other case corresponding to $\Gamma_{B/X}/M_{B/X}=0.01$,
we present accessible discovery and exclusion mass limits on VLB/VLX
for different production processes and systematic uncertainty cases
are shown in Table \ref{tab:Limit-table-1}.

\begin{table}
\caption{Mass limit for evidence ($3\sigma$) / discovery ($5\sigma$) and
exclusion ($2\sigma$) of VLB/VLX signal at the HL-LHC with $L_{int}=3$
ab$^{-1}$. Here, the relative width is taken as $\Gamma_{B/X}/m_{B/X}=0.1$.
Each panel corresponds to different systematic uncertainty case, denoting
no systematics, a mild systematic case $\delta_{sys}=10\%$ and a
possible systematic case $\delta_{sys}=20\%$. \label{tab:Limit-table}}

\begin{tabular}{|c|c|c|}
\hline 
Process & \multicolumn{2}{c|}{Mass Limits {[}GeV{]}}\tabularnewline
\hline 
\hline 
$\delta_{sys}=0$ & Evidence ($3\sigma$) / Discovery ($5\sigma$) & Exclusion ($2\sigma$)\tabularnewline
\hline 
Bbj & 2418 / 2273 & 2491\tabularnewline
\hline 
Btj & 2018 / 1764 & 2145\tabularnewline
\hline 
Xtj & 2273 / 2145 & 2364\tabularnewline
\hline 
$\delta_{sys}=10\%$ &  & \tabularnewline
\hline 
Bbj & 2344 / 2163 & 2490\tabularnewline
\hline 
Btj & 1909 / 1672 & 2144\tabularnewline
\hline 
Xtj & 2235 / 2055 & 2345\tabularnewline
\hline 
$\delta_{sys}=20\%$ &  & \tabularnewline
\hline 
Bbj & 2400 / 2236 & 2490\tabularnewline
\hline 
Btj & 1964 / 1655 & 2144\tabularnewline
\hline 
Xtj & 2273 / 2109 & 2344\tabularnewline
\hline 
\end{tabular}
\end{table}

\begin{table}
\caption{The same as Table \ref{tab:Limit-table}, but for relative width $\Gamma_{B/X}/m_{B/X}=0.01$.
\label{tab:Limit-table-1}}

\begin{tabular}{|c|c|c|}
\hline 
Process & \multicolumn{2}{c|}{Mass Limits {[}GeV{]}}\tabularnewline
\hline 
\hline 
$\delta_{sys}=0$ & Evidence ($3\sigma$) / Discovery ($5\sigma$) & Exclusion ($2\sigma$)\tabularnewline
\hline 
Bbj & 1927 / 1764 & 2018\tabularnewline
\hline 
Btj & 1382 / 1273 & 1455\tabularnewline
\hline 
Xtj & 1800 / 1618 & 1873\tabularnewline
\hline 
$\delta_{sys}=10\%$ &  & \tabularnewline
\hline 
Bbj & 1926 / 1745 & 2017\tabularnewline
\hline 
Btj & 1345 / 1255 & 1436\tabularnewline
\hline 
Xtj & 1782 / 1600 & 1872\tabularnewline
\hline 
$\delta_{sys}=20\%$ &  & \tabularnewline
\hline 
Bbj & 1891 / 1655 & 2000\tabularnewline
\hline 
Btj & 1309 / 1218 & 1400\tabularnewline
\hline 
Xtj & 1745 / 1491 & 1872\tabularnewline
\hline 
\end{tabular}
\end{table}

\section{Conclusions}

\label{sec:conclusions-1} 

We have investigated single production of the VLB quarks in the 4FNS
scheme at 14 TeV HL-LHC via the process $pp\to Bbj$ and $pp\to Btj/Xtj$
with a subsequent decay channel $B\to Wt(t\to Wb)$ (where the $W$
bosons decay hadronically $W\to jj$) in a simplified model framework
of vector-like quarks B/X, where only two parameters are contained
in one type of VLQs, the VLB/VLX quark mass $m_{B/X}$ and the decay
width-to-mass ratio ($\Gamma_{B/X}/m_{B/X}$). Then, our study is
relevant to the VLB in one of the singlet ($B$), doublet (T B) or
triplet (T B Y) branching scenarios for the $Bbj$ and $Btj$ production,
while it is relevant to the VLX in one of the doublet (X T) or triplet
(X T B) scenarios for the $Xtj$ production. We have performed a fast
detector simulation for the signal and the relevant SM backgrounds.
This search presents a significant advance over previous searches
for VLB/VLX searches at the pp collisions. We present our results
showing that, with an integrated luminosity of 3 ab$^{-1}$ at the
HL-LHC, the discovery range can reach the VLB (VLX) mass up to $2273(2145)$
GeV without considering the systematic uncertainty. When a possible
systematic uncertainty of $20\%$ is included, we find an accesible
range of mass up to $2236(2109)$ GeV, respectively. On the other
hand, the 95\% CL exclusion limits for the masses up to $2490(2344)$
GeV for VLB(VLX), in the case of a possible systematic uncertainty
of $\delta_{sys}=20\%$. We see that the excluding capabilities (a
similar interpretation for the discovery capability) of VLB/VLX searches
are enhanced with the increase of both the center-of-mass energy and
integrated luminosity of HL-LHC. We expect our study can be guideline
search for a possible VLB/VLX quark at the future pp colliders.

\section*{Acknowledgements}

\label{sec:acknowledgements-1} 

We would like to thank Ankara University High Energy Physics (AUHEP)
group for fruitful discussions. The numerical calculations reported
in this paper were partially performed at TUBITAK ULAKBIM, High Performance
and Grid Computing Center (TRUBA resources). The work was supported
in part by the Turkish Energy, Nuclear and Mineral Research Agency
(TENMAK) under project grant No. 2020TAEK(CERN)A5.H1.F5-25.

\end{document}